\documentclass[a4paper,11pt]{article}
\usepackage{graphicx,rotating,slashed,amsmath,xcolor,amssymb,amsfonts,colortbl,cite,subcaption,mathtools}
\usepackage[totalheight = 23cm, totalwidth = 17cm]{geometry}
\usepackage{physics}
\usepackage[makeroom]{cancel}
\usepackage{tikz-feynman}
\usepackage{jheppub}

\let\latexd\d
\definecolor{bluc}{cmyk}{1,1,0,0.1}
\definecolor{rossoCP3}{cmyk}{0,.88,.77,.40}
\definecolor{rosso}{cmyk}{0,1,1,0.4}
\definecolor{rossos}{cmyk}{0,1,1,0.55}
\definecolor{rossoc}{cmyk}{0,1,1,0.2}
\definecolor{verdes}{cmyk}{0.92,0,0.59,0.4}
\definecolor{violet}{rgb}{0.58, 0.0, 0.83}

\hypersetup{colorlinks, bookmarksopen, bookmarksnumbered,
citecolor=verdes, linkcolor=blue, pdfstartview=FitH, urlcolor=blue}

\def\a{\alpha}        
\def\b{\beta}       
\def\g{\gamma}    
\def\d{\delta}    
\def\e{\epsilon}

\def\z{\zeta}

\def\na{\nabla}
\def\bo{\Box}
\def\nn{\nonumber}
\def\d{\delta}

\def\pa{\partial} \def\del{\partial}

\def\uno{\mbox{1 \kern-.59em {\rm l}}}

\title{Extending Weinberg's EFT: effective scalar-tensor theories up to sixth order}

\author[a]{Eugeny Babichev}
\author[b]{Suk\latexd{r}ti Bansal}
\author[c,d]{Maria Mylova}
\author[e,f]{Antonio Padilla}

\affiliation[a]{Universit\'e Paris-Saclay, CNRS/IN2P3, IJCLab, 91405 Orsay, France} 
\affiliation[b]{Institute for Theoretical Physics, TU Wien, Wiedner Hauptstraße 8-10/136, A-1040 Vienna, Austria} 
\affiliation[c]{Kavli Institute for the Physics and Mathematics of the Universe (WPI),
The University of Tokyo, Kashiwa, Chiba 277-8583, Japan} 
\affiliation[d]{Perimeter Institute for Theoretical Physics, Waterloo, Ontario N2L 2Y5, Canada} 
\affiliation[e]{School of Physics and Astronomy, University of Nottingham, Nottingham NG7 2RD, UK} 

\affiliation[f]{Nottingham Centre of Gravity, University of Nottingham, Nottingham NG7 2RD, UK}

\emailAdd{ eugeny.babichev@ijclab.in2p3.fr}
\emailAdd{sukrti.b@gmail.com}
\emailAdd{maria.mylova@ipmu.jp}
\emailAdd{antonio.padilla@nottingham.ac.uk}

\abstract{We present a systematic construction of the six-derivative  effective scalar-tensor theories, extending the four-derivative framework previously developed  by Steven Weinberg. The on-shell effective field theory comprises five parity-even and three parity-odd independent six-derivative scalar-tensor interactions, representing all inequivalent deformations consistent with general covariance. We further confirm this operator counting through an independent analysis using the scattering amplitude formalism in four-dimensional flat spacetime. The six-derivative Lagrangian constructed here provides the next-to-leading-order extension of scalar-tensor gravity, furnishing a robust framework for exploring quantum or stringy corrections, parity-violating interactions, and strong-curvature effects in cosmology, black hole physics and gravitational wave observations.}

\begin{document} 

\maketitle

\section{Introduction}
\label{sec:intro}

Over the past two decades, modified gravity \cite{Clifton:2011jh,Joyce:2014kja} has attracted sustained interest, initially driven by cosmological applications such as inflation \cite{Weinberg:2008hq} and dark energy \cite{Copeland:2006wr}, and later invigorated by the detection of gravitational waves \cite{LIGOScientific:2016aoc,LIGOScientific:2017vwq} and the prospect of precision tests of gravity across a wide range of scales \cite{Berti:2005ys, Berti:2015itd}. In parallel, substantial advances in numerical relativity have enabled detailed explorations of the dynamics of modified theories in strongly gravitating regimes, including black-hole binaries \cite{Pretorius:2005gq}.  

A common and conceptually simple route to modifying gravity is to introduce an additional scalar degree of freedom that mediates the gravitational interaction alongside the graviton. While early formulations of scalar--tensor theories focused on the Brans--Dicke framework \cite{Brans:1961sx} and its generalisations, more recent work has centred on galileons \cite{Nicolis:2008in}, Horndeski \cite{Horndeski:1974wa,Deffayet:2011gz}  and beyond Horndeski theories \cite{Gleyzes:2014dya,Goldberger:2004jt}. Horndeski theories are often celebrated as the most general scalar--tensor models with manifestly second-order field equations, and have therefore formed the backbone of numerous studies in cosmology, strong-gravity physics, and gravitational-wave phenomenology. The restriction to second-order equations is typically motivated by the desire to avoid Ostrogradsky instabilities associated with higher derivatives \cite{Ostrogradsky:1850fid}.

From the perspective of effective field theory (EFT), however, the community's fixation on maintaining manifestly second-order field equations is somewhat misplaced. As emphasised in \cite{Burgess:2014lwa}, higher-derivative operators generically arise in EFTs when heavy fields are integrated out order-by-order in a derivative expansion. A simple illustration appears in the EFT of the Goldstone mode of a spontaneously broken scalar: integrating out the heavy radial (Higgs) mode induces operators that lead to higher-order equations of motion, yet the resulting theory is perfectly healthy. Within the regime of validity of the EFT, the derivative expansion is controlled, couplings remain weak, and no instabilities are triggered by the higher derivatives.  

Once modified gravity is viewed through the EFT lens, the question of identifying the ``most general'' scalar--tensor theory must be reformulated. Allowing higher-derivative operators removes the special status of Horndeski and beyond Horndeski (including Degenerate Higher-Order Scalar-Tensor theories~\cite{Langlois:2015cwa}), and in principle seems to open the door to an overwhelmingly large operator space. Nevertheless, this space is far from unmanageable. Many operators that appear distinct at first sight are related through structural identities of the underlying geometry -- for example, algebraic and differential identities of the Riemann tensor  -- as well as by integrations by parts \cite{Henning:2015daa,Solomon:2017nlh,Ruhdorfer:2019qmk,Daas:2024pxs}.  

Our goal is to construct an independent and minimal operator basis for scalar--tensor theories, organised order-by-order in a derivative expansion. The basis at each order is not unique: integration by parts and identities allow one to map between different choices. What matters is that \emph{some} minimal basis exists, and that the resulting Lagrangians capture the full dynamics of the \emph{most general} scalar--tensor theory at that order consistent with the symmetries of the underlying theory and available degrees of freedom and away from any external sources. The coupling to external sources is not invariant under general field redefinitions, and we leave a systematic treatment of this issue to future work.

Weinberg famously presented the scalar–tensor EFT including all terms up to fourth order in derivatives \cite{Weinberg:2008hq}.  The minimal basis at fourth order contains three operators: two with even parity, one with odd.  While Weinberg’s original motivation was to apply this to inflation, more recent work has applied the framework to black hole binaries and other strong-gravity regimes \cite{AresteSalo:2022hua}. 

The goal of the present work is to extend Weinberg's formalism to sixth order in derivatives. Extending the scalar--tensor EFT to include six-derivative terms is well motivated. Higher-derivative operators arise generically in effective descriptions of gravity that capture quantum or string-theoretic corrections to the Einstein--Hilbert action. In principle, the coefficients of these operators encode information about the UV completion, via matching. At fourth order in derivatives, the EFT operators include curvature-squared invariants, known to appear in low-energy string expansions  \cite{Zwiebach:1985uq,Zumino:1985dp,Cecotti:1985nf} (see also \cite{Campbell:1990fu,Cunillera:2021fbc}). Six-derivative operators constitute the next-to-next leading order, including cubic curvature terms such as \(R_{\alpha\beta\epsilon\zeta} R^{\alpha\beta\gamma\delta} R_{\gamma\delta}{}^{\epsilon\zeta}\) and mixed scalar--curvature structures like \(R_{\beta\gamma\delta\epsilon} R^{\beta\gamma\delta\epsilon} \nabla_{\alpha}\phi\,\nabla^{\alpha}\phi\) or \(R_{\alpha\gamma\beta\delta}\,\nabla^{\alpha}\phi\,\nabla^{\beta}\phi\,\nabla^{\delta}\nabla^{\gamma}\phi\).

Perhaps surprisingly, six-derivative  terms can  sometimes dominate over four-derivatives, even at low energies. In pure gravity this is automatic: in four dimensions the Gauss--Bonnet (and Pontryagin) terms are topological, so the first  non-trivial higher-derivative corrections
really are the cubic-curvature terms at six-derivatives.  
Although scalar--tensor theories allow non-trivial four-derivative interactions, scenarios with suppressed scalar dependence at fourth order could shift the leading corrections to sixth order. This may be especially important in inflation where scalar potentials are expected to be relatively flat\footnote{As an example, consider a theory with the following interactions: $\alpha e^{\epsilon \phi} \big(R_{\alpha\beta\gamma\delta}R^{\alpha\beta\gamma\delta}
    - 4 R_{\alpha\beta}R^{\alpha\beta}
    + R^2\big)+\frac{1}{\Lambda^2}R_{\alpha\beta\gamma\delta} R^{\alpha\beta}{}_{\zeta\eta} R^{\gamma\delta\zeta\eta}$ where $\Lambda$ is the cut-off controlling the expansion in derivatives and $\alpha$ is some dimensionless order one coupling.  The parameter $\epsilon$ sets a flatness scale in the $\phi$ dependent coupling. When we compute the field equations, the contribution of the fourth order term is suppressed by $\epsilon$. Thus, if we take $\epsilon$ to be sufficiently small, the sixth order term will dominate.}. 

As we have stated above, the final form of the scalar-tensor EFT is not unique, up to any given order in the derivative expansion.  However, a minimal representation of the most general Lorentz-invariant theory is given as follows,
\begin{equation}
    S=S_\text{LO}[g, \phi] +S_4[g, \phi]+S_6[g, \phi]+\ldots \, .
\end{equation}
Here, the leading order piece is just the Einstein-Hilbert action alongside a canonical scalar with a potential, 
\begin{equation}
    S_\text{LO}[g, \phi]=\int \dd^4 x \sqrt{-g}\,\bigg[\frac{M_P^2}{2} R-\frac{M^2}{2}\,\na_\a\phi\,\na^\a\phi-M_P^2\,U(\phi)\bigg]\,. \nn
\end{equation} 
The fourth-order corrections are most elegantly presented in terms of Horndeski operators and the Pontryagin operator, 
\begin{multline}
S_4[g, \phi]
    = \int \dd^4 x \sqrt{-g}\,\bigg[ f_1(\phi)\,(\nabla_\alpha\phi\,\nabla^\alpha\phi)^2
    + f_9(\phi)\,\big(R_{\alpha\beta\gamma\delta}R^{\alpha\beta\gamma\delta}
    - 4 R_{\alpha\beta}R^{\alpha\beta}
    + R^2\big) \\
    + f_{10}(\phi)\,\epsilon^{\alpha\beta\gamma\delta}
    R_{\alpha\beta}{}^{\zeta\eta}R_{\gamma\delta\zeta\eta} \Bigg]\,,\nn
\end{multline}
while the six-derivative corrections obtained in this work take the form
\begin{small}
\begin{multline}
S_6[g, \phi]
    = \int \dd^4 x \sqrt{-g}\,\Bigg[ 
e_1(\phi)R_{\alpha\beta\gamma\delta} R^{\alpha\beta}{}_{\zeta\eta} R^{\gamma\delta\zeta\eta}
+e_{16}(\phi)R_{\alpha\beta\gamma\delta} R^{\alpha\beta\gamma\delta} \nabla_{\zeta}\phi \nabla^{\zeta}\phi \\+e_{22}(\phi)R_{\alpha\beta\gamma\delta} \nabla^{\alpha}\phi \nabla^{\gamma}\phi \nabla^{\delta}\nabla^{\beta}\phi\, \nn+e_{67}(\phi) (\nabla_\alpha \phi \, \nabla^\alpha \phi)^3 +e_{71}(\phi) \nabla_\alpha \phi \, \nabla^\alpha \phi \, \nabla_\gamma \nabla_\beta \phi \, \nabla^\gamma \nabla^\beta \phi\\+o_1(\phi)\,\epsilon_{\zeta \eta \theta \kappa}R_{\alpha \beta \gamma \delta}  R^{\alpha \beta\zeta \eta} R^{\gamma \delta\theta \kappa}
+o_7(\phi)\,\epsilon_{\gamma\delta\zeta\eta}R_{\alpha\beta}{}^{\gamma\delta} R^{\alpha\beta\zeta\eta}  \nabla_{\theta} \phi \nabla^{\theta} \phi+o_{13}(\phi)\,\epsilon_{\gamma \delta\zeta \eta} R_{\alpha \beta}{}^{\gamma \delta}\,\nabla^{\alpha} \phi \nabla^{\zeta} \phi \nabla^{\eta} \nabla^{\beta} \phi
\Bigg]\,.
\end{multline}
\end{small}
This expression is our main result. Note that this expression can equivalently be presented in the so-called Weyl basis, simply by replacing each Riemann tensor with the corresponding Weyl tensor. The $\phi$ dependent functions at fourth order, $f_n(\phi)$, are dimensionless, while at sixth order, $e_n(\phi)$ and $o_n(\phi)$, they scale like $1/\Lambda^2$ where $\Lambda$ is the cut-off in the derivative expansion.

Six-derivative scalar–tensor EFTs have been explored previously, and explicit Lagrangians have been presented for models exhibiting shift symmetry in the scalar field \cite{Solomon:2017nlh, Ruhdorfer:2019qmk}, albeit with different results, with \cite{Ruhdorfer:2019qmk} arguing that \cite{Solomon:2017nlh} missed eliminating certain redundant terms via integration by parts. Indeed, in the shift-symmetric limit, our result agrees with \cite{Ruhdorfer:2019qmk}. Since none of the six-derivative operators are topological in 4D, imposing shift symmetry on our terms reduces their coefficient functions to constants, reproducing the operator basis of \cite{Ruhdorfer:2019qmk}. However, our result goes beyond shift symmetry. We also include operators of both odd and even parity. The latter can include chirality in gravitational waves, which affects the amplitude and velocity of left- and right-handed tensor modes, and in the case of a pseudo-scalar coupling they can also manifest as cosmic birefringence, i.e. a rotation of the polarisation of CMB photons. 

There are well-motivated scenarios (for instance with approximate shift symmetry) in which four-derivative odd-operators are parametrically suppressed, so that the leading higher-derivative corrections effectively start at six-derivatives. This has been a topic of active research with several papers studying the three or higher-point parity-violating tensor statistics of the CMB~\cite{Satoh:2010ep, Maldacena:2011nz, Shiraishi:2011st, Soda:2011am, Shiraishi:2016mok, Crisostomi:2017ugk, Bartolo:2020gsh, Bordin:2020eui, Gong:2023kpe, Moretti:2024fzb}.

The remainder of this paper is organized as follows. In Section \ref{sec:onshell} we review how to construct a minimal on-shell operator basis for scalar--tensor EFTs. In Section \ref{4derEFT}, we briefly recall Weinberg's four-derivative scalar--tensor EFT.  Section \ref{OnShellEFT} presents the construction of the six-derivative scalar–tensor EFT, distinguishing between even- and odd-parity sectors, and detailing the standard EFT procedure for identifying independent operators and eliminating redundancies. The resulting on-shell six-derivative Lagrangian is then assembled in Subsection \ref{6derEFT}. Section \ref{scattering-amp} outlines a counting of
the on-shell EFT terms up to sixth order using on-shell scattering amplitudes, providing
an alternative and compact representation of our results. We conclude in Section \ref{summary} with a discussion of the implications of our results and possible extensions of this framework. Appendix \ref{identities} shows the identities used in this work while Appendix \ref{offshell_app} contains the complete covariant enumeration of six-derivative terms before imposing the equations of motion or using integrations by parts to remove redundant operators.

Throughout this paper, we have made extensive use of the Mathematica package xAct to help perform our calculations.

\section{Eliminating redundancies in scalar--tensor theories} \label{sec:onshell}

Although the formalism of actions and Lagrangians is a powerful tool for writing down effective theories in physics, it is littered with redundancies. One often begins by enumerating all possible operators up to a given mass dimension or number of derivatives, but many of these operators are not independent. Integration by parts can relate superficially distinct terms, and identities that hold only in specific spacetime dimensions can further collapse the operator basis. The latter are particularly important in geometric theories of gravity, where Bianchi, Ricci and dimension-dependent tensor identities — such as those involving the Riemann tensor or the antisymmetrisation of indices — can make certain higher-curvature combinations redundant or topological. As a result, different Lagrangians, though formally distinct, can yield identical local physics when expressed in terms of  observables.

Additionally, in a local EFT, one can perform perturbative field redefinitions without changing the analytic structure of the asymptotic states, so the S-matrix remains invariant. This freedom allows one to shift unwanted operators to higher orders in the perturbative expansion, rendering them redundant.

At the level of classical dynamics, these field redefinitions leave the structure of vacuum solutions unchanged, since the equations of motion are equivalent under invertible transformations. However, their effect can become nontrivial when sources or boundary terms are included --- an issue that deserves more detailed examination, and to which we will return in future work.

To understand how this works, consider  a generic EFT whose dynamical fields we collectively denote by $\varphi$, with indices suppressed for simplicity. We organize the action as a derivative expansion
\begin{align}
S[\varphi] = S_\text{LO}[\varphi]+S_4[\varphi]+ S_6[\varphi] + \ldots \, ,
\label{eq:01}
\end{align}
where $S_\text{LO}[\varphi]\equiv S_0[\varphi]+S_2[\varphi]$ contains the leading order terms in the EFT, including operators with at most two derivatives. In general, $S_{2N}[\varphi]$ contain operators with exactly $2N$ derivatives. We now consider the following field redefinition, order by order in derivatives:
\begin{equation}
\begin{split}
\varphi \rightarrow \varphi + \mathcal{O}_2[\varphi]+\mathcal{O}_4[\varphi] +\ldots \,.
\label{eq:02}
\end{split}
\end{equation}
As before,  the operators $\mathcal{O}_{2N}[\varphi]$ contain exactly $2N$ derivatives. Under this field redefinition, 
$S[\varphi] \rightarrow \tilde{S}[\varphi] = S[\varphi]  + \Delta S$, with a functional Taylor expansion giving
\begin{equation}\begin{split}
\Delta S = \sum_{n \geq 1} \frac{1}{n!} \int_{x_1} \dd^4{x}_1 \cdots \int_{x_n} \dd^4{x}_n \frac{\delta^n S}{\delta \varphi(x_1) \cdots \delta \varphi(x_n)} \sum_{m_1, \ldots, m_n \geq 1}\mathcal{O}_{2m_1}[\varphi](x_1) \cdots \mathcal{O}_{2m_n}[\varphi](x_n) \,.
\label{eq:03}
\end{split}\end{equation}
It follows that
\begin{multline}
   \Delta S_{2N}= \sum_{n \geq 1} \frac{1}{n!} \int_{x_1} \dd^4{x}_1 \cdots \int_{x_n} \dd^4{x}_n \sum_{M \geq 0} \frac{\delta^n S_{2M}}{\delta \varphi(x_1) \cdots \delta \varphi(x_n)} \\ \times \sum_{m_1, \ldots, m_n \geq 1}\mathcal{O}_{2m_1}[\varphi](x_1) \cdots \mathcal{O}_{2m_n}[\varphi](x_n) \delta_{m_1+\ldots +m_n, N-M} \,.
\end{multline}
Since the field redefinition has been expanded order by order in derivatives, with no change at zeroth order, we find that $\Delta S_0=\Delta S_2=0$. For $N \geq 2$, however, we obtain
\begin{equation}
    \Delta S_{2N} \supset  \int \dd^4 x \frac{\delta S_0}{\delta \varphi(x)}\mathcal{O}_{2N}[\varphi](x)+\int \dd^4 x \frac{\delta S_2}{\delta \varphi(x)}\mathcal{O}_{2(N-1)}[\varphi](x) \, .
\end{equation}
If $S_{2N}$ contains a term proportional to the leading-order equation of motion, $\delta S_2 / \delta \varphi$, such a term can be removed by an appropriate field redefinition. 
Specifically, if 
\[
S_{2N} \supset \int \! \dd^4 x \, 
\frac{\delta S_2}{\delta \varphi(x)} \,
\mathcal{F}_{2(N-1)}[\varphi](x) \, ,
\]
then choosing
\[
\mathcal{O}_{2(N-1)}[\varphi](x)
= - \mathcal{F}_{2(N-1)}[\varphi](x)\, ,
\]
eliminates this contribution at order $2N$.

This operation, however, induces new terms to-next-to-leading order in the derivative expansion.
In particular, it generates one {\it lower}-order term of the form
\[
-\int \! \dd^4 x \,
\frac{\delta S_0}{\delta \varphi(x)} \,
\mathcal{F}_{2(N-1)}[\varphi](x).
\]

This reflects the standard Wilsonian EFT lore: once the theory is truncated at a given order, one can use field redefinitions to eliminate operators proportional to the equations of motion of the quadratic theory and treat them as redundant. In our example, this corresponds to the elimination of the highest-order terms proportional to $\delta S_2 / \delta \varphi$, and is sometimes described as going from the \emph{off-shell} EFT to  the \emph{on-shell} EFT \cite{Georgi:1991ch}. 
$(\delta S_2 / \delta \varphi)\,(\ldots)$ 
with lower-order in derivatives terms 
$-(\delta S_0 / \delta \varphi)\,(\ldots)$, 
which is equivalent to imposing the leading-order equations of motion, $\delta S_\text{LO} / \delta \varphi \equiv \delta S_0 / \delta \varphi+\delta S_2 / \delta \varphi=0$ within the Lagrangian.

Let us re-iterate that the resulting minimal basis of the EFT is not unique, since different field redefinitions lead to physically equivalent descriptions. In this work we follow the standard convention of eliminating operators proportional to the equations of motion.

In this paper, we focus on scalar-tensor theories with vanishing torsion, where the leading order EFT is given by the Einstein-Hilbert action alongside a canonical scalar field:
\begin{equation} \label{SLO}
    S_\text{LO}[g, \phi]=\int \dd^4 x \sqrt{-g}\,\bigg[\frac{M_P^2}{2} R-\frac{M^2}{2}\,\na_\a\phi\,\na^\a\phi-M_P^2\,U(\phi)\bigg] \, .
\end{equation}
Here $M_P=1/\sqrt{8\pi G}$ is the reduced Planck mass, $M$ is the generic mass scale and $U(\phi)$ is the potential for a dimensionless scalar $\phi=\varphi/M$, with $\varphi$ being a general scalar field with $[\varphi]=\,\,$mass. The leading-order equations of motion read,
\begin{align}\label{eom}
    \frac{\d{\cal S}_\text{LO}}{\d \phi}
&=   \frac{\d{\cal S}_0}{\d \phi} +\frac{\d{\cal S}_2}{\d \phi}  = \sqrt{-g}\, M^2\,\bo\phi - \sqrt{-g}\, M_P^2\, U'(\phi) = 0 \, ,\nn 
\\
    \frac{\d{\cal S}_\text{LO}}{\d g^{\a\b}} & = \frac{\d{\cal S}_0}{\d g^{\a\b}}+\frac{\d{\cal S}_2}{\d g^{\a\b}}  =\sqrt{-g}\, \frac{M_P^2}{2} \left(R_{\a\b} -\frac12 R\, g_{\a\b}\right) -\sqrt{-g}\, \frac{M^2}{2}\left(\nabla_\alpha \phi \nabla_\beta \phi-\frac12\, g_{\alpha \beta} \,(\nabla \phi)^2\right)  \nn \\
    &\qquad\qquad\qquad\qquad\,\,-\sqrt{-g}\,\frac{M^2}{2}\,U(\phi)\, g_{\a\b} =0 \, .
\end{align}
In practice, we will use the scalar field equations to eliminate terms proportional to   $\Box \phi$  and the metric field equations to eliminate terms proportional to the Ricci tensor as well as the Ricci scalar. 

Our generic algorithm for constructing the most general scalar–tensor theory up to a given order in derivatives, modulo redundancies, is as follows.
We first list all operators in the derivative expansion consistent with diffeomorphism invariance.
For a given order of derivatives, all the operators can be classified by the number of Riemann tensors (including Ricci tensor and Ricci scalar), and by the
number of the scalar fields $\phi$ which appear with at least one derivative acting on them (e.g., $\nabla\phi$, $\nabla\nabla\phi$ etc.), which we will refer to as \textit{dressed} scalars  for convenience. 
We denote each such set of operators by $\mathcal{S}_r^j$, where $r$ refers to the number of Riemann tensors in the operator and $j$ is the number of dressed scalars.

As it has been discussed above, there are a number of tools to eliminate redundant terms: symmetries of Riemann and Levi-Civita tensor, Bianchi identities (BI), Dimensional Dependent Identities (DDIs), Ricci identities, integration by parts (IBP) and leading order equations of motion (\ref{eom}). 
Note that applying symmetries, Bianchi identities and Dimensional Dependent Identities  do not change the number of Riemann tensors, nor the number of dressed $\phi$'s, i.e., 
$$
\mathcal{S}_r^j \big|_{\text{BI, DDI, symmetries}} \to \mathcal{S}_r^j \, . 
$$
On the other hand, the remaining operations can change both the number of curvature tensors $r$ and the number of dressed scalars $j$, appearing in any given  term $\mathcal{S}_r^j$. 
Indeed, the use of Ricci identities introduces an extra Riemann, e.g., 
$$
\mathcal{S}_r^j\big|_{\text{Ricci identities}} \to \mathcal{S}_r^j + \mathcal{S}_{r+1}^j \, . 
$$
An integration by parts potentially introduces an extra derivative of a scalar, i.e., 
$$
\mathcal{S}_r^j \big|_{\text{IBP}} \to \mathcal{S}_r^j + \mathcal{S}_{r}^{j+1} \, . 
$$
Finally, using the scalar field equations of motion (\ref{eom}), removes one dressed scalar from $S^j$, effectively mapping such term to $S^{j-1}$, so the corresponding operators are redundant and can be discarded at this order.

\begin{figure}[t]
\begin{center}
\includegraphics[width=0.9\textwidth]{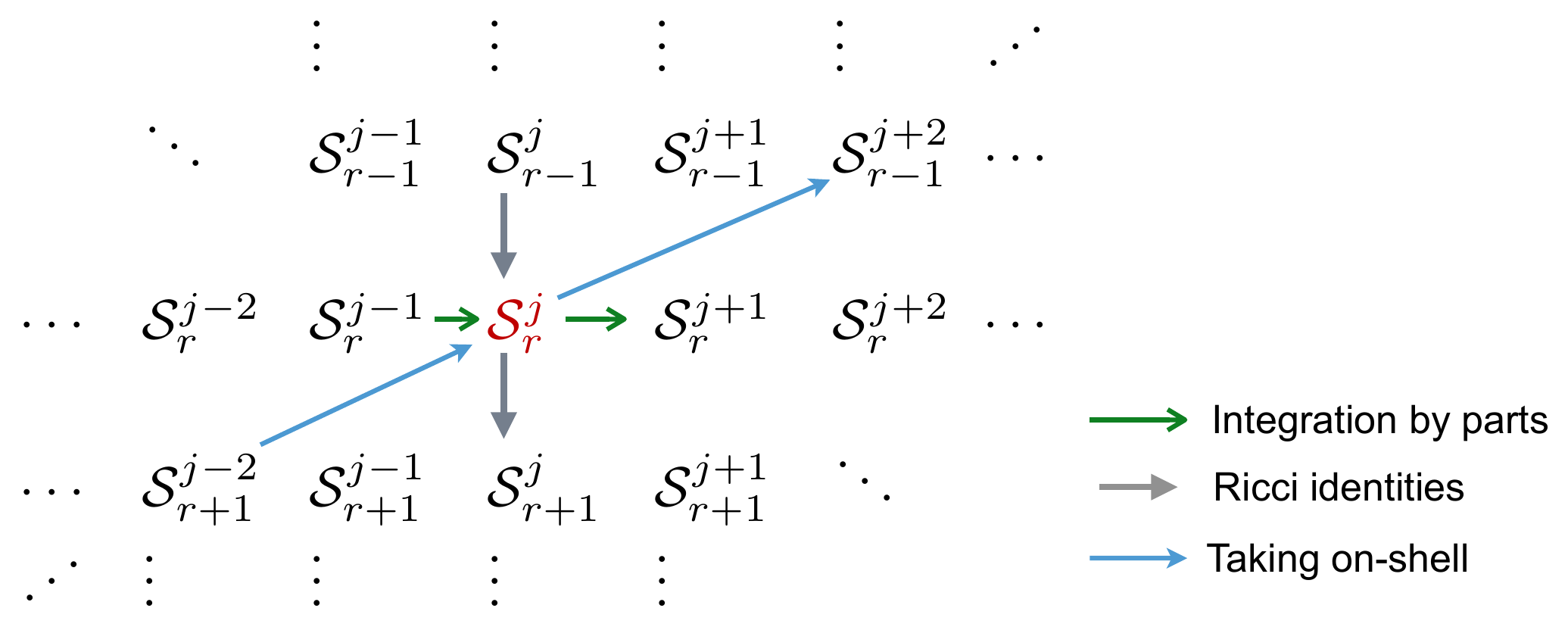}
\caption{The diagram illustrates how a typical set of terms $\mathcal{S}_r^j$ transforms under the application of Bianchi identities, dimensional-dependent identities, Ricci identities, integration by parts, and the leading-order equations of motion (\ref{eom}).
The use of symmetries, Bianchi identities, and dimensional-dependent identities does not generate terms belonging to other groups.
In contrast, Ricci identities, integration by parts, and the leading-order equations of motion (\ref{eom}) may produce terms outside the original set, as indicated by the arrows in the diagram.
}
\label{diagram}
\end{center}
\end{figure}

When the metric equations in (\ref{eom}) are  applied, we trade one Riemann (in the form of Ricci tensor or Ricci scalar) for $\sim \nabla\phi\nabla\phi$ plus lower-derivative terms that we dismiss (these can be absorbed into a redefinition of the potential), i.e., we have
$$
\mathcal{S}_r^j \big|_{\text{EOMs}} \to  \mathcal{S}_{r-1}^{j+2} \, . 
$$
Fig.~\ref{diagram} illustrates the transitions affecting a typical set of terms $\mathcal{S}_r^j$ under various operations.
As the ``flow'' under all the operations is defined in only one direction and the diagram contains no ``loops'', each set $\mathcal{S}_r^j$ for each $r$ and $j$ can be considered independently, without any risk of losing terms. This procedure relies on the requirement that we do not move opposite to the direction of the arrows. In particular, operations such as replacing the Riemann tensor by a commutator of derivatives acting on $\phi$ are not permitted.

It is a general procedure that applies for construction non-redundant Lagrangian with any number of derivatives.

\subsection{Weinberg's four-derivative 
EFT}\label{4derEFT}

We now review the derivation of the general scalar-tensor theory up to fourth order in derivatives, as originally discussed in the context of inflation \cite{Weinberg:2008hq}. As we have just described, the first task is to write down all possible operators up to this order. This gives $S=S_\text{LO}[g, \phi]+S_4[g, \phi]$ where the leading order action is given by \eqref{SLO} and $S_4[g, \phi]=\int \dd^4 x \sqrt{-g} \,\mathcal{L}_4$. The four-derivative \textit{off-shell} scalar-tensor Lagrangian is given by \cite{Weinberg:2008hq}\footnote{Although the Lagrangian in \cite{Weinberg:2008hq} is written in terms of the Weyl tensor instead of the Riemann tensor, it can be checked that it is equivalent to Lagrangian \eqref{4derLagOffShell}.}
\begin{align}\label{4derLagOffShell}
    {\cal L}_4=&f_1(\phi)\,(\na_\a\phi\na^\a\phi)^2+f_2(\phi)\na_\a\phi\na^\a\phi\,\bo\phi+f_3(\phi)(\bo\phi)^2+f_4(\phi)R_{\a\b}\,\na^\a\phi\na^\b\phi\nn\\
    &+f_5(\phi)R\,\na_\a\phi\na^\a\phi+f_6(\phi)R\,\bo\phi+f_7(\phi)R^2+f_8(\phi)R_{\a\b}R^{\a\b}+f_9(\phi)\,R_{\a\b\g\d}R^{\a\b\g\d}\nn\\
    &+f_{10}(\phi)\,\epsilon^{\a\b\g\d} R_{\a\b}{}^{\z\eta}R_{\g\d\z\eta}\,.
\end{align}
Here $f_n(\phi)$ are arbitrary functions of the scalar field and $\epsilon^{\a\b\g\d}$ is the four-dimensional Levi-Civita tensor. Next, we substitute the leading order equations from (\ref{eom})  to replace $\Box \phi$, the Ricci tensor and the Ricci scalar. In this {\it on-shell} limit, the above Lagrangian becomes
\begin{align}\label{4derLag}
    {\cal L}_4=f_1(\phi)\,(\na_\a\phi\na^\a\phi)^2+f_9(\phi)\,R_{\a\b\g\d}R^{\a\b\g\d}+f_{10}(\phi)\,\epsilon^{\a\b\g\d} R_{\a\b}{}^{\z\eta}R_{\g\d\z\eta}\,.
\end{align}
This expression captures the most general scalar–tensor theory at fourth order in derivatives that respect diffeomorphism invariance. Any other operator at the same order can be obtained through integration by parts, or the use of identities and therefore does not introduce new physics. The result contains three independent operators, each multiplied by an arbitrary function of the scalar field. These operators can be regarded as a convenient basis for the space of fourth-order scalar–tensor interactions.

The choice of basis for the on-shell EFT is not unique. For example, we could replace the Riemann tensors with Weyl tensors in  \eqref{4derLag}, recovering the fourth order on-shell EFT presented by Weinberg \cite{Weinberg:2008hq},
\begin{align}\label{4derLag0}
    {\cal L}_4=f_1(\phi)\,(\na_\a\phi\na^\a\phi)^2+f_9(\phi)\,C_{\a\b\g\d}C^{\a\b\g\d}+f_{10}(\phi)\,\epsilon^{\a\b\g\d} C_{\a\b}{}^{\z\eta}C_{\g\d\z\eta}\,.
\end{align}
Another elegant choice replaces the square of the Riemann tensor with the Gauss–Bonnet invariant, yielding
\begin{align}\label{4derLag1}
    {\cal L}_4
    = f_1(\phi)\,(\nabla_\alpha\phi\,\nabla^\alpha\phi)^2
    + f_9(\phi)\,\big(R_{\alpha\beta\gamma\delta}R^{\alpha\beta\gamma\delta}
    - 4 R_{\alpha\beta}R^{\alpha\beta}
    + R^2\big)
    + f_{10}(\phi)\,\epsilon^{\alpha\beta\gamma\delta}
    R_{\alpha\beta}{}^{\zeta\eta}R_{\gamma\delta\zeta\eta}\,.
\end{align}

Of course, at any given order, the number of operators in the a given  basis is fixed. At fourth order, we see that the basis must include two operators of even parity and one of odd.  Any scalar tensor EFT containing additional operators at this order, must contain redundancies. 

\subsection{Extending Weinberg's EFT: on-shell scalar-tensor theories up to sixth order in derivatives}\label{OnShellEFT}
We now go beyond the four-derivative EFT derived by Weinberg to sixth order in derivatives. Since the number of operators  is somewhat larger, we split our analysis in to parity even and parity odd operators. 
\subsection{Even parity terms}\label{even_onshell}
First we describe an algorithm for efficiently generating all possible interaction operators at sixth order. In particular, all six-derivative terms of even parity without a derivative of the scalar field $\phi$ can be generated by first evaluating
\begin{align}\label{Even6DerRiem}
\na^{\a_1}(\na^{\a_2}(\na^{\a_3}(\na^{\a_4}R_{\b_1\b_2\b_3\b_4}))) \, ,
\end{align}
and then contracting the indices of the resulting expressions in all possible ways to form scalars.\footnote{If the contractions are performed before fully expanding/evaluating the nested derivatives, many admissible six-derivative terms fail to appear.} To construct all even-parity six-derivative terms involving derivatives of $\phi$, one instead computes
\begin{align}\label{Even6DerPhi}
\na^{\a_1}(\na^{\a_2}(\na^{\a_3}(\na^{\a_4}(\na^{\a_5}(\na^{\a_6}\phi^6))))) \, ,
\end{align}
and contracts the resulting indices in every admissible way to obtain scalars. Since commuting covariant derivatives generates Riemann tensors, \eqref{Even6DerPhi} systematically produces all possible six-derivative terms involving derivatives of $\phi$.

Alternatively, one may first generate all five-derivative terms by the same procedure, contracting all but one index, and then apply a total derivative to obtain the complete set of six-derivative structures. Expressing all six-derivative terms as originating from total derivatives of five-derivative ones enables efficient elimination of redundancies via integration by parts.

The next step in our algorithm is to discard all the terms rendered redundant by the identities in appendix \ref{identities}. This leaves us with 89 independent six-derivative terms of even parity, presented in equation \eqref{89_even} of appendix \ref{offshell_app}. The off-shell terms are shown in eq. \eqref{even_offshell}. We now take the Lagrangian on-shell. This involves substituting eqs. \eqref{eom} to replace $\Box \phi$, the Ricci tensor, and the Ricci scalar, by their expressions in terms of $U(\phi)$ and single derivatives of $\phi$. Identities in appendix \ref{identities} are reapplied at this stage to prevent redundant terms from reappearing. The result leaves us with fourteen independent operators,
\begin{small}
\begin{align}\label{OnShellEvenBefIBP}
&e_1(\phi)R_{\alpha\beta\gamma\delta} R^{\alpha\beta}{}_{\zeta\eta} R^{\gamma\delta\zeta\eta}
,&&\!\!\!\!e_2(\phi)\nabla_\zeta R_{\alpha\beta\gamma\delta} \, \nabla^\zeta R^{\alpha\beta\gamma\delta}
,&&\!\!\!\!e_{16}(\phi)R_{\alpha\beta\gamma\delta} R^{\alpha\beta\gamma\delta} \nabla_{\zeta}\phi \nabla^{\zeta}\phi,\nn\\
&e_{18}(\phi)R^{\alpha\beta\gamma\delta} \nabla_{\zeta} R_{\alpha\beta\gamma\delta} \nabla^{\zeta}\phi
,&&\!\!\!\!e_{22}(\phi)R_{\alpha\beta\gamma\delta} \nabla^{\alpha}\phi \nabla^{\gamma}\phi \nabla^{\delta}\nabla^{\beta}\phi,&&\!\!\!\!e_{23}(\phi)R_{\alpha\beta\gamma\delta} \nabla^{\gamma}\nabla^{\alpha}\phi \nabla^{\delta}\nabla^{\beta}\phi, \nn\\
&e_{67}(\phi) (\nabla_\alpha \phi \, \nabla^\alpha \phi)^3 ,&&\!\!\!\!e_{69}(\phi)\nabla_\alpha \phi \, \nabla^\alpha \phi \, \nabla_\beta \phi\,\nabla_\gamma\phi \, \nabla^\gamma \nabla^\beta \phi,&&\!\!\!\!e_{71}(\phi) \nabla_\alpha \phi \, \nabla^\alpha \phi \, \nabla_\gamma \nabla_\beta \phi \, \nabla^\gamma \nabla^\beta \phi,\nn\\
&e_{72}(\phi)\nabla^\alpha \phi \, \nabla^\beta \phi \, \nabla_\gamma \nabla_\beta \phi \, \nabla^\gamma \nabla_\alpha \phi,&&\!\!\!\!e_{75}(\phi)\nabla_\alpha \phi \, \nabla_\beta \phi \, \nabla_\gamma \phi \,\nabla^\gamma \nabla^\beta \nabla^\alpha \phi ,&&\!\!\!\!e_{78}(\phi)\nabla_{\beta}\nabla_{\alpha}\phi\,\nabla_{\gamma}\nabla^{\beta}\phi\,\nabla^{\gamma}\nabla^{\alpha}\phi,\nn\\
&e_{81}(\phi)\nabla_\alpha \phi \,\nabla_\gamma \nabla_\beta \phi\, \nabla^\gamma \nabla^\beta \nabla^\alpha \phi ,&&\!\!\!\!e_{85}(\phi)\nabla_{\gamma}\nabla_{\beta}\nabla_{\alpha}\phi\,\nabla^{\gamma}\nabla^{\beta}\nabla^{\alpha}\phi.
\end{align}
\end{small}
Here $e_n(\phi)=\tilde e_n(\phi)/\Lambda^2$ where $\tilde e_n(\phi)$ are arbitrary dimensionless algebraic functions of $\phi$ and $\Lambda$ is a scaling parameter with $[\Lambda]=$ mass.

At first glance, this appears to complete the construction of the on-shell EFT. However, some redundancies persist, with several of the fourteen operators equivalent under integration by parts. Removing this leaves five independent terms. If a previously eliminated off-shell term reappears during the procedure, it is removed again by substituting its on-shell expression. The flow described above  automatically chooses the basis with derivatives distributed as evenly as possible, giving the following set:
\begin{align}\label{OnshellEven}
&e_1(\phi)R_{\alpha\beta\gamma\delta} R^{\alpha\beta}{}_{\zeta\eta} R^{\gamma\delta\zeta\eta}
\,,&&e_{16}(\phi)R_{\alpha\beta\gamma\delta} R^{\alpha\beta\gamma\delta} \nabla_{\zeta}\phi \nabla^{\zeta}\phi\,,&&e_{22}(\phi)R_{\alpha\beta\gamma\delta} \nabla^{\alpha}\phi \nabla^{\gamma}\phi \nabla^{\delta}\nabla^{\beta}\phi\,, \nn\\
&e_{67}(\phi) (\nabla_\alpha \phi \, \nabla^\alpha \phi)^3 \,,&&e_{71}(\phi) \nabla_\alpha \phi \, \nabla^\alpha \phi \, \nabla_\gamma \nabla_\beta \phi \, \nabla^\gamma \nabla^\beta \phi\,.
\end{align}
While the terms above were selected to minimize the number of derivatives acting on a scalar and tensor, one may in general choose a different basis. By performing integration by parts, we can select, from the set of terms in eq. \eqref{OnShellEvenBefIBP}, any one of the two terms without a scalar, any two of the four terms consisting of both the Riemann tensor as well as the scalar field, and any two of the eight scalar-only terms.

\subsection{Odd parity terms}\label{odd_onshell}

All six-derivative terms of odd parity without derivatives of the scalar $\phi$, arise by computing
\begin{align}\label{Odd6DerRiem}
    \e_{\a_1\a_2\a_3\a_4}\na^{\b_1}(\na^{\b_2}(\na^{\b_3}(\na^{\b_4}R_{\g_1\g_2\g_3\g_4}))) \, ,
\end{align}
and contracting the indices of the resulting expressions into scalars. Similarly, those consisting of $\phi$-derivatives are generated from
\begin{align}\label{Odd6DerPhi}
\e_{\a_1\a_2\a_3\a_4}\na^{\b_1}(\na^{\b_2}(\na^{\b_3}(\na^{\b_4}(\na^{\b_5}(\na^{\b_6}\phi^6)))))\,.
\end{align}
Redundant terms are subsequently eliminated through the identities listed in appendix \ref{identities}. This gives a total of sixteen independent 6-derivative terms of odd parity (cf. \eqref{16_odd} in appendix \ref{offshell_app}). This number is significantly smaller than the corresponding even-parity terms (89 in total), due to the additional antisymmetry-symmetry patterns introduced by the Levi-Civita tensor, which introduce greater redundancy. The off-shell terms are given in eq. \eqref{odd_offshell}. Upon going on-shell as described in section \ref{even_onshell}, the sixteen terms reduce further to the following six terms, with coefficients $o_n(\phi)=\tilde o_n(\phi)/\Lambda^2$ where $\tilde o_n(\phi)$ are general algebraic functions of $\phi$.
\begin{small}
\begin{align}\label{OnShellOddBefIBP}
&o_1(\phi)\,\epsilon_{\zeta \eta \theta \kappa}R_{\alpha \beta \gamma \delta}  R^{\alpha \beta\zeta \eta} R^{\gamma \delta\theta \kappa},&&\!\!\!\!o_2(\phi)\,\epsilon_{\gamma \delta \zeta\eta}\nabla_{\theta} R_{\alpha \beta}{}^{\gamma \delta}\nabla^{\theta} R^{\alpha \beta\zeta\eta} ,&&\!\!\!\!o_7(\phi)\,\epsilon_{\gamma\delta\zeta\eta}R_{\alpha\beta}{}^{\gamma\delta} R^{\alpha\beta\zeta\eta}  \nabla_{\theta} \phi \nabla^{\theta} \phi,\nn\\
&o_{12}(\phi)\,\epsilon_{\gamma \delta\zeta \eta} R_{\alpha \beta}{}^{\gamma \delta} \nabla^{\beta} R^{\alpha \zeta}\nabla^{\eta} \phi,&&\!\!\!\!o_{13}(\phi)\,\epsilon_{\gamma \delta\zeta \eta} R_{\alpha \beta}{}^{\gamma \delta}\,\nabla^{\alpha} \phi \nabla^{\zeta} \phi \nabla^{\eta} \nabla^{\beta} \phi,&&\!\!\!\!o_{14}(\phi)\,\epsilon_{\gamma \delta\zeta \eta} R_{\alpha \beta}{}^{\gamma \delta} \nabla^{\zeta} \nabla^{\alpha} \phi \nabla^{\eta} \nabla^{\beta} \phi.
\end{align}\end{small}Upon eliminating redundant terms via integration by parts, three independent terms remain. Choosing terms with lowest possible derivatives on a scalar/tensor, we get the following three terms:
\begin{small}
\begin{align}\label{OnshellOdd}
&o_1(\phi)\,\epsilon_{\zeta \eta \theta \kappa}R_{\alpha \beta \gamma \delta}  R^{\alpha \beta\zeta \eta} R^{\gamma \delta\theta \kappa},&&\!\!\!\!o_7(\phi)\,\epsilon_{\gamma\delta\zeta\eta}R_{\alpha\beta}{}^{\gamma\delta} R^{\alpha\beta\zeta\eta}  \nabla_{\theta} \phi \nabla^{\theta} \phi,&&\!\!\!\!o_{13}(\phi)\,\epsilon_{\gamma \delta\zeta \eta} R_{\alpha \beta}{}^{\gamma \delta}\,\nabla^{\alpha} \phi \nabla^{\zeta} \phi \nabla^{\eta} \nabla^{\beta} \phi \, .
\end{align}
\end{small}In general, integrating by parts without preferring a specific basis can give us any one of the two terms without a scalar in eq. \eqref{OnShellOddBefIBP}, and any two of the four terms consisting of both the Riemann tensor as well as the scalar field in eq. \eqref{OnShellOddBefIBP}. 

\subsection{Six-Derivative EFT}\label{6derEFT}
We now bring together the results from sections \ref{even_onshell} and \ref{odd_onshell} to express the general  on-shell scalar tensor effective theory up to sixth order in derivatives. This is given by $S=S_{LO}[g, \phi]+S_4[g, \phi]+S_6[g, \phi]$ where the leading order action is given by \eqref{SLO} and $S_{2n}=\int \dd^4 x \mathcal{L}_{2n}$. The on-shell EFT at fourth order is given by \eqref{4derLag}, whereas at sixth order we have,  
\begin{align}\label{6der_onshell}
    {\cal L}_6=\,\,& e_1(\phi)R_{\alpha\beta\gamma\delta} R^{\alpha\beta}{}_{\zeta\eta} R^{\gamma\delta\zeta\eta}
+e_{16}(\phi)R_{\alpha\beta\gamma\delta} R^{\alpha\beta\gamma\delta} \nabla_{\zeta}\phi \nabla^{\zeta}\phi+e_{22}(\phi)R_{\alpha\beta\gamma\delta} \nabla^{\alpha}\phi \nabla^{\gamma}\phi \nabla^{\delta}\nabla^{\beta}\phi \nn\\
&+e_{67}(\phi) (\nabla_\alpha \phi \, \nabla^\alpha \phi)^3 +e_{71}(\phi) \nabla_\alpha \phi \, \nabla^\alpha \phi \, \nabla_\gamma \nabla_\beta \phi \, \nabla^\gamma \nabla^\beta \phi+o_1(\phi)\,\epsilon_{\zeta \eta \theta \kappa}R_{\alpha \beta \gamma \delta}  R^{\alpha \beta\zeta \eta} R^{\gamma \delta\theta \kappa}\nn\\
&+o_7(\phi)\,\epsilon_{\gamma\delta\zeta\eta}R_{\alpha\beta}{}^{\gamma\delta} R^{\alpha\beta\zeta\eta}  \nabla_{\theta} \phi \nabla^{\theta} \phi+o_{13}(\phi)\,\epsilon_{\gamma \delta\zeta \eta} R_{\alpha \beta}{}^{\gamma \delta}\,\nabla^{\alpha} \phi \nabla^{\zeta} \phi \nabla^{\eta} \nabla^{\beta} \phi\,.
\end{align}
This set of operators with their $\phi$ dependent coefficients provide a basis for EFT at this order in derivatives. As explained at fourth order, this basis is not unique. We could perform integration by parts,  use identities or carry out field redefinitions, generating a new basis of operators. For example, we could consider a new basis to the one given in \eqref{6der_onshell}, trading each Riemann tensor for a Weyl tensor --- this would differ from \eqref{6der_onshell} by terms proportional to the Ricci tensor and therefore be equivalent. Of course, recall that the number of operators in the basis is fixed at any given order. At sixth  order the  basis includes  five operators  of even parity and three odd.  If you have a theory with more independent operators at this order, there must be redundancies.

\section{On-shell scattering amplitudes}\label{scattering-amp}

Our aim in this paper is to classify all independent scalar–tensor interactions up to sixth order in derivatives. By {\it independent}, we mean interactions that are inequivalent under field redefinitions, integration by parts and identities. On-shell scattering amplitudes in Minkowski space are automatically invariant under local field redefinitions and  encode integration-by-parts identities through energy–momentum conservation. They therefore provide an ideal framework for checking our analysis of the previous section. Since we ultimately wish to match these amplitudes to operators in the corresponding EFT, it is sufficient to focus on amplitude contributions without propagator poles — that is, on local contact interactions in the EFT.

Consider, for example,  an interaction operator of the form $c\,\sqrt{-g}\, R_{\mu\nu\alpha \beta}\, \nabla^\mu \phi \nabla^\alpha \phi \nabla ^\nu \nabla^\beta \phi$. Identifying the graviton as the metric fluctuation on a Minkowski background, $g_{\mu\nu}=\eta_{\mu\nu} +h_{\mu\nu}$, such an interaction will  contribute to the following contact diagram 
\begin{center}
\begin{tikzpicture}
  \begin{feynman}
    \vertex (v) at (0,0);

    \vertex (h) at (-1.2,1.2) {$h_{\mu\nu}$};
    \vertex (p1) at (1.2,1.2) {$\phi$};
    \vertex (p2) at (-1.2,-1.2) {$\phi$};
    \vertex (p3) at (1.2,-1.2) {$\phi$};

    \diagram*{
      (v) -- [photon] (h),     
      (v) -- [plain] (p1),     
      (v) -- [plain] (p2),
      (v) -- [plain] (p3),
    };
  \end{feynman}
\end{tikzpicture}
\end{center}
along with higher-point contact diagrams with multiple gravitons,  of the form
\begin{center}
\begin{tikzpicture}
  \begin{feynman}
    \vertex (v) at (0,0);

    \vertex (sNE) at ( 1.4,  1.4);
    \vertex (sSE) at ( 1.4, -1.4);
    \vertex (sSW) at (-1.4, -1.4);

    \vertex (gTop)    at (115:1.55);
    \vertex (gMid)    at (140:1.55);
    \vertex (gBottom) at (165:1.55);

    \diagram*{
      (v) -- [plain]  (sNE),
      (v) -- [plain]  (sSE),
      (v) -- [plain]  (sSW),
      (v) -- [photon] (gTop),
      (v) -- [photon] (gMid),
      (v) -- [photon] (gBottom),
    };
  \end{feynman}

  \def\R{1.65}
  \def\Astart{162}  
  \def\Aend{118}    

  \draw[dotted, thick] (0,0) ++(\Astart:\R)
    arc[start angle=\Astart, end angle=\Aend, radius=\R];
\end{tikzpicture}
\end{center}
due to the non-linear nature of gravity. Although this operator contributes to contact diagrams with arbitrarily many external gravitons, it is sufficient to consider the four-point diagram with a single graviton to identify its presence, at least schematically.   Of course, any on-shell {\it amplitude} will receive contributions from both contact and exchange diagrams, with the latter characterised by poles. To read off the full structure of the interaction, we need to go beyond four-point amplitudes, where diffeomorphism invariance, through the amplitude Ward identities, fully constrains the higher order structure. However, in this section, we are only really interested in the schematic structure, so the analytic part of the four-point amplitude should be enough. 

We can actually say a bit more. Each operator should really be understood to appear with a general coefficient $c(\phi)$. In this case we are really considering
\[
c(\phi)\sqrt{-g}\, R_{\mu\nu\alpha\beta}\, \nabla^\mu \phi\, \nabla^\alpha \phi\, \nabla^\nu \nabla^\beta \phi \, .
\]
Expanding the coefficient as
\[
c(\phi) = c_0 + c_1 \phi + c_2 \phi^2 + \cdots ,
\]
the four-point contact diagram corresponds to the $c_0$ term, while the contributions from $c_p$ with $p = 1, 2, \ldots$ simply add $p$ external $\phi$ legs to the same diagram, without introducing additional derivatives. In the soft limit, where the momenta on these extra scalar legs vanish, evaluation of the resulting $(4+p)$-point diagram reduces to an evaluation of the underlying four-point diagram. Thus,  the analytic part of the four-point amplitude already suffices to infer the existence of interaction of the more general form $
c(\phi)\sqrt{-g}\, R_{\mu\nu\alpha\beta}\, \nabla^\mu \phi\, \nabla^\alpha \phi\, \nabla^\nu \nabla^\beta \phi$, at least schematically. 

Our aim in this section is to use on-shell amplitudes to  establish which independent interactions schematically appear at a given order in derivatives. In principle, we might worry that the presence of dimensionless scalars and gravitons means we need to consider an arbitrarily large number of diagrams to capture all the possible interactions that might appear. However, the analysis of the previous paragraph suggests this is not the case. Before we get stuck in to the detailed structure of the amplitudes, let us ask which diagrams we need to consider to schematically capture all the independent operators at a given order $n$ in derivatives. 

Consider a generic operator, schematically of the form $\del^{n_1} (\text{Riemann tensor})^{n_2} \phi^{n_3}$. This has a total of $n=n_1+2n_2$ derivatives and contributes  to amplitudes with  $n_2+\delta n_2$ gravitons and $n_3$ scalars, where $\delta n_2 \geq 0$. In others words, it contributes to $N=n_2+\delta n_2+n_3$ point  contact diagrams of the form,
\begin{center}
\begin{tikzpicture}
  \begin{feynman}
    \vertex (v) at (0,0);
    \vertex (g1) at (-1.6, 1.2);
    \vertex (g2) at (-1.6, 0.4);
    \vertex (g3) at (-1.6,-0.4);
    \vertex (g4) at (-1.6,-1.2);

    \vertex (s1) at ( 1.6,  1.4);
    \vertex (s2) at ( 1.6,  0.7);
    \vertex (s3) at ( 1.6,  0.0);
    \vertex (s4) at ( 1.6, -0.7);
    \vertex (s5) at ( 1.6, -1.4);
    \diagram*{
      (v) -- [photon] (g1),
      (v) -- [photon] (g2),
      (v) -- [photon] (g3),
      (v) -- [photon] (g4),

      (v) -- [plain] (s1),
      (v) -- [plain] (s2),
      (v) -- [plain] (s3),
      (v) -- [plain] (s4),
      (v) -- [plain] (s5),
    };
  \end{feynman}
  \draw[dotted, thick]
    (-1.8,1.3) .. controls (-2.0,0.0) .. (-1.8,-1.3)
    node[midway, left, xshift=-3pt] {$n_2+\delta n_2$};
  \draw[dotted, thick]
    (1.8,-1.5) .. controls (2.0,0.0) .. (1.8,1.5)
    node[midway, right, xshift=3pt] {$n_3$};
\end{tikzpicture}
\end{center}
As in the example above, the $\delta n_2$ encodes the fact that at a given order $n$ in derivatives a single interaction operator contributes to an infinite number of higher-point diagrams with an arbitrarily large number of external gravitons. As above, to infer the presence of the operator,  it is enough to set $\delta n_2=0$ and equate the number of external gravitons to the number of Riemann tensor terms.  

The schematic form of the operator $\del^{n_1} (\text{Riemann tensor})^{n_2} \phi^{n_3}$ suppresses all the possible different distributions of derivatives. To capture them all, what contact diagrams do we need to consider? For a given order in derivatives $n$, we need to make sure we include enough contact diagrams to reproduce every allowed interaction structure. This requires allowing the derivatives to be shared evenly across the curvature tensors and the scalar fields. In other words, as long as we consider contact diagrams for which
$n_2+n_3 \geq n_1$,
we ensure that all relevant interactions are represented without introducing redundant cases.  Since we also assumed that $\delta n_2=0$, this implies that it is always enough to consider contact diagrams for which the number of external legs is greater than or equal to the number of explicit derivatives in the interaction operator, $N>n_1$.

To systematically work through the operators and the corresponding contact diagrams, we fix the number of derivatives $n$ and the number of Riemann tensors in the interaction, $n_2$.  To recover all possible interaction structures, we have seen that we need to work with diagrams with $N \geq n_1=n-2n_2$ external legs.  For $n=4$ it follows that $N \geq 4-2n_2$. Since the number of Riemann tensors is always non-negative $n_2 \geq 0$, it is enough to focus on four-point interactions.  For $n=6$, it follows that $N \geq 6-2n_2$.  For pure scalar interactions with no Riemann tensors, we need to consider up to six-point interactions to capture all the independent operators. When Riemann tensors are present, it is enough to just consider four-point interactions. 

Let us start by studying in detail amplitudes that capture the effects of operators up to fourth order in derivatives. As we have just seen, it is enough to consider four-point contact diagrams, corresponding to the four-point amplitude contribution without propagator poles.  We begin with scalar four-point amplitudes. We label each of the external legs with four-momentum $p_i$ for $i=1,2,3,4$ and correspondingly introduce the standard Mandelstam variables
\begin{eqnarray}
    s &=&(p_1+p_2)^2=(p_3+p_4)^2 \, ,\\
    t &=&(p_1+p_3)^2=(p_2+p_4)^2 \, ,\\
    u &=&(p_1+p_4)^2=(p_2+p_3)^2 \, .
\end{eqnarray}
The amplitude must be invariant under particle exchange $p_i \leftrightarrow p_j$, which is equivalent to exchanging Mandelstam variables. Thus the amplitude must be given in terms of symmetric polynomials in $s$, $t$ and $u$. At fourth order in derivatives and in the absence of poles, this gives a general four-point scalar amplitude, 
\begin{equation}
    \mathcal{A}^{(4)}_{\phi \phi \phi \phi}(1^0, 2^0, 3^0, 4^0)=\alpha (s+t+u)^2+\beta (st +tu+us) \, ,
\end{equation}
for arbitrary constants $\alpha $ and $\beta$. Now it is well known that $(s+t+u)=-4m^2$ where $m$ is the mass of the scalar. If this combination appears, it just lowers the number of derivatives allowing us, without loss of generality, to set $m=0$ for simplicity.  For the scalar four-point amplitude, this leaves us with
\begin{equation}
    \mathcal{A}^{(4)}_{\phi \phi \phi \phi}(1^0, 2^0, 3^0, 4^0)=\beta (st +tu+us) \, ,
\end{equation}
corresponding to the standard K-essence operator $\sqrt{-g} (\nabla \phi)^4$. This points to a more general interaction of the form $\sqrt{-g}\, c(\phi)(\nabla \phi)^4$. 

Staying with fourth order in derivatives, we now introduce gravitons into the external legs of our four-point diagrams.  To do this efficiently, we need to implement the spinor-helicity formalism.  The spinor-helicity formalism has become a cornerstone of modern amplitude theory, providing a compact and symmetry-transparent framework for describing scattering processes in four-dimensional flat spacetime. By expressing null momenta as bispinors \( p_{\alpha \dot{\alpha}} = \lambda_\alpha \tilde{\lambda}_{\dot{\alpha}} \), external states can be represented directly in terms of helicity spinors, making Lorentz covariance and little-group transformations manifest. On-shell amplitudes for the scattering of massless particles can then be systematically constructed at tree level from a small set of simple rules. For pedagogical introductions to the formalism and its applications, see \cite{Elvang:2013cua,Elvang:2015rqa,Henn:2014yza,Dixon:2011xs}.

Let us briefly review some of the basic ingredients of the formalism.  We have already noted that for a massless particle, the momentum can be written as
$
    p_{\alpha \dot{\alpha}} = \lambda_\alpha \tilde{\lambda}_{\dot{\alpha}} ,
$
where $\lambda_\alpha$ and $\tilde{\lambda}_{\dot{\alpha}}$ are commuting two-component spinors of opposite chirality. The freedom to rescale these spinors as
\begin{equation}
    \lambda \;\to\; t\,\lambda \, , \qquad
    \tilde{\lambda} \;\to\; t^{-1} \tilde{\lambda}\, ,
\end{equation}
leaves $p_{\alpha\dot{\alpha}}$ invariant.  This rescaling corresponds to the \emph{little group} for massless particles — physically, a \(U(1)\) phase rotation that acts on the particle’s polarization.  
A particle of helicity \(h\) picks up a phase \(t^{-2h}\) under this transformation. Hence, a generic on-shell amplitude transforms under little-group scaling as
\begin{equation}
    \mathcal{A}(1^{h_1}, 2^{h_2}, \ldots)
    \;\to\;
    t_1^{-2h_1} t_2^{-2h_2} \cdots \mathcal{A} \, .
\end{equation}
For scalars $h=0$, while for gravitons $h=\pm 2$.

We define the spinor contractions
\begin{equation}
    \langle ij \rangle = \epsilon^{\alpha\beta} \lambda_{i,\alpha}\lambda_{j,\beta} \, , \qquad 
    [ij] = \epsilon^{\dot{\alpha}\dot{\beta}} \tilde{\lambda}_{i,\dot{\alpha}}\tilde{\lambda}_{j,\dot{\beta}} \, .
\end{equation}
Each bracket carries a single power of the little-group parameter:
\begin{equation}
    \langle ij \rangle \sim t_i t_j \, , \qquad [ij] \sim t_i^{-1} t_j^{-1} \, .
\end{equation}
Thus, every positive-helicity leg ($h>0$) must appear with enough $[\,\cdot\,]$ brackets
and every negative-helicity leg ($h<0$) with enough $\langle\,\cdot\,\rangle$ brackets to yield $t^{-2h}$ scaling. The basic rules for constructing our scalar--tensor amplitudes, without propagator poles,  are as follows: 
\begin{enumerate}
\item Each $\langle ij\rangle$ or $[ij]$ has units of momentum.
    \item Brackets are antisymmetric:
        $\langle ij\rangle = - \langle ji\rangle, \quad [ij] = -[ji].
$    
    \item Each $h=+2$ graviton contributes four angled brackets; each $h=-2$ graviton contributes four square brackets.
    \item Scalars come with no net brackets (equal numbers of $\langle\,\cdot\,\rangle$ and $[\,\cdot\,]$).
    \item There are no brackets in the denominator. This is required to avoid unwanted poles in our four-point amplitudes.
    \item Momentum conservation is equivalent to $ \sum_i i] \langle i=0$
    \item In four dimensions, the Schouten identity (valid for any spinors $i,j,k,l$) gives 
        $ [ij][kl] - [ik][jl] - [il][kj] = 0$,  and similarly for angle brackets.
\end{enumerate}
We now consider a four-point diagram with one graviton (labelled with a 1) and three scalars (labelled with 2,3 and 4). For a graviton with positive helicity, we need the amplitude to at least include $[1 \cdot ][1 \cdot ][1 \cdot ][1 \cdot ] $. To get the right scaling for the scalars, we must include at least four compensating angled brackets. As there are no poles, this would yield at least eight powers of momentum, which is too many, given we are working up to fourth order in derivatives. We conclude that the corresponding amplitude should vanish $ \mathcal{A}^{(4)}_{h \phi \phi \phi}(1^{+2}, 2^0, 3^0, 4^0)=0$. Similar considerations imply that the amplitude vanishes up to fourth order for a negative helicity graviton, $ \mathcal{A}^{(4)}_{h \phi \phi \phi}(1^{-2}, 2^0, 3^0, 4^0)=0$. This is consistent with the fact that operators of the form $\phi\, G_{\mu\nu} \nabla^\mu \phi \nabla^\nu \phi$ can be eliminated on account of redundancy.

For four-point diagrams with two gravitons and two scalar, a similar analysis yields the following results for the amplitude up to fourth order
\begin{eqnarray}
    \mathcal{A}^{(4)}_{h h \phi \phi}(1^{+2}, 2^{+2}, 3^0, 4^0) &=&\alpha_{++} [12]^4 \, ,\\  \mathcal{A}^{(4)}_{h h \phi \phi}(1^{+2}, 2^{-2}, 3^0, 4^0) &=&0 \, , \\ \mathcal{A}^{(4)}_{h h \phi \phi}(1^{-2}, 2^{-2}, 3^0, 4^0) &=&\alpha_{--} \langle 12\rangle ^4 \, ,
\end{eqnarray}
for constants $\alpha_{++}$ and $\alpha_{--}$. When $\alpha_{++}=\alpha_{--}$, the corresponding operator is parity even; when $\alpha_{++}=-\alpha_{--}$, the corresponding operator is parity odd. At fourth order in derivatives, this suggests one operator of each parity of the form $c(\phi) (\text{Riemann tensor})^2$. For even parity, this corresponds to the Gauss-Bonnet operator, $\sqrt{-g}\, \phi^2 (R_{\mu\nu\alpha\beta}R^{\mu\nu\alpha\beta}-4 R_{\mu\nu}R^{\mu\nu}+R^2)$. For odd parity, it corresponds to the Pontryagin operator, $\sqrt{-g}\,  \phi^2 
\epsilon^{\a\b\g\d} R_{\a\b}{}^{\z\eta}R_{\g\d\z\eta}
$. In each case, this points to the same operator with $\phi^2$ replaced with a general function of $\phi$. 

There are no operators with more than two Riemann tensors at fourth order in derivatives. To infer the schematic form of our interactions, we argued that it was enough to set the number of external gravitons to the number of Riemann tensor terms. We therefore, do not need to consider more than two gravitons in our amplitudes at fourth order in derivatives in order to schematically infer all the interactions of interest. 

We now turn our attention to amplitudes at sixth order in derivatives. Once again, we start with pure scalar interactions, which at this order requires us to consider six point interactions.  As the six-point scalar diagram has six external legs, we introduce four-momentum $p_i$ for $i=1, \ldots, 6$ and a generalisations of the Mandelstam variables, $s_{ij}=p_i \cdot p_j$. Of course, these variables contain several redundancies from both energy conservation and the on-shell conditions.  As before, because the amplitude is invariant under particle exchange, it should be built from symmetric polynomials in these variables. At sixth order in derivatives, taking into account the redundancies, there are just two different contributions to the six-point scalar amplitude,
\begin{equation}
    \mathcal{A}^{(6)}_{\phi \phi\phi \phi \phi \phi}(1^{0}, 2^{0}, 3^0, 4^0, 5^0, 6^0)=\gamma (s_{12} s_{23} s_{34} +\text{perms}) +\delta (s_{12}s_{34}s_{56}+\text{perms})  \, ,
\end{equation}
corresponding to the following operators 
$$\sqrt{-g}\, \phi^2\, \nabla_\mu \phi \nabla^{\mu}{}\nabla_\nu \phi \nabla^\mu \nabla_\alpha \phi \nabla^\alpha \phi \, , \qquad  \sqrt{-g}\, (\nabla \phi)^6 \, .$$ As usual, this points to the same  operators weighted by arbitrary functions of $\phi$.

We now include gravitons. At sixth order in derivatives, the presence of one or more gravitons means it is enough to consider four-point amplitudes again.  For one graviton and three scalars, we find the following amplitudes
\begin{eqnarray}
    \mathcal{A}^{(6)}_{h \phi \phi \phi}(1^{+2}, 2^{0}, 3^0, 4^0) &=&\beta_{+} ([12]^2[13][14]\langle 23\rangle \langle 24 \rangle +\text{perms})\, ,\\ 
 \mathcal{A}^{(6)}_{h \phi \phi \phi}(1^{-2}, 2^{0}, 3^0, 4^0) &=&\beta_{-} (\langle 12\rangle ^2\langle 13\rangle \langle 14\rangle [ 23][ 24 ] +\text{perms}) \, ,
\end{eqnarray}
for arbitrary constants $\beta_\pm$. As before, $\beta_+=\pm \beta_-$ maps to parity even (odd) operators of the form $\del^4 (\text{Riemann tensor} ) \phi^3$. Again, we can generalise to the same operators weighted by arbitrary functions of $\phi$.  

For two gravitons and two scalars, we naively obtain the following amplitudes at sixth order
\begin{eqnarray}
    \mathcal{A}^{(6)}_{h h \phi \phi}(1^{+2}, 2^{+2}, 3^0, 4^0) &=&\gamma_{++} ([12]^4 s+\text{perms})+\delta_{++}([12]^3[13][24]\langle 34 \rangle+\text{perms}) \, , \\  
    \mathcal{A}^{(6)}_{h h \phi \phi}(1^{+2}, 2^{-2}, 3^0, 4^0) &=&0 \, , \\ \mathcal{A}^{(6)}_{h h \phi \phi}(1^{-2}, 2^{-2}, 3^0, 4^0) &=&\gamma_{--} (\langle 12\rangle ^4s +\text{perms})+\delta_{--}(\langle 12\rangle ^3\langle 13\rangle \langle 24\rangle [ 34 ]+\text{perms}) \, ,
\end{eqnarray}
for arbitrary constants.  However, upon use of the Schouten identity, we can show that the two non-trivial  terms are equivalent in each of the amplitudes. Since the Schouten identity is a feature of four dimensions, this is reminiscent of applying a dimensionally dependent identity. The equivalence between terms implies that there is just one parity even operator of the form $\del^4 (\text{Riemann tensor} )^2 \phi^2$, and one parity odd. We can, of course,  generalise to the same operators weighted by arbitrary functions of $\phi$.

Finally, we consider operators with three Riemann tensors, which can be appear at sixth order in derivatives. This requires us to look at  amplitudes with three gravitons and one scalar.  The only non-vanishing contributions are as follows
\begin{eqnarray}
    \mathcal{A}^{(6)}_{h h h \phi}(1^{+2}, 2^{+2}, 3^{+2}, 4^0) &=&\lambda_{+++} ([12]^2[23]^3[31]^2+\text{perms})\\  
\mathcal{A}^{(6)}_{h h h \phi}(1^{-2}, 2^{-2}, 3^{-2}, 4^0) &=&\lambda_{---} (\langle 12\rangle^2\langle 23\rangle^2 \langle 31\rangle^2 +\text{perms})
\end{eqnarray}
for arbitrary constants $\lambda_{\pm \pm \pm}$. As before, $\lambda_{+++}=\pm \lambda_{---}$ maps to a parity even (odd) operator of the form $(\text{Riemann tensor} )^3 \phi$, and we can generalise to the same operators weighted by arbitrary functions of $\phi$. At sixth order in derivatives there are no terms with more than three Riemann tensors, so we are done.

Let us summarize what our amplitude analysis has revealed.  At fourth order in derivatives, we can describe the EFT in terms of the following independent operators:
\begin{itemize}
    \item even parity: one with no Riemann tensors and one with two Riemann tensors,
    \item odd parity:  one with two Riemann tensors.
\end{itemize}
At sixth order, we can describe the EFT in terms of the following: 
\begin{itemize}
    \item even parity: two with no Riemann tensors, one each of one Riemann tensor, two Riemann tensors and three Riemann tensors,
    \item odd parity:  one each of one Riemann tensor, two Riemann tensors and three Riemann tensors.
\end{itemize}
This counting agrees perfectly with our analysis of the previous sections. The on-shell amplitude methods therefore provide an independent and efficient check of those results. If we disregard the pure-scalar sector and focus only on operators containing at least one Riemann tensor, there is a one-to-one correspondence between even- and odd-parity operators. This follows from the fact that all mixed-helicity amplitudes vanish at this order in derivatives. Should this persist at higher derivative order, the same correspondence would continue to hold. Whether or not this happens remains to be checked explicitly. It is important to note that different choices of operator basis in the EFT may redistribute derivatives and Riemann tensors among operators, but they cannot change the total number of independent interactions. In contrast, the amplitudes capture the genuinely invariant content of the theory, independent of basis or field redefinition.
	
\section{Summary and Outlook}\label{summary}

In this work we have systematically constructed the on-shell six-derivative scalar-tensor EFT, extending the four-derivative framework of \cite{Weinberg:2008hq} to next-to-leading order. Our classification exhausts all independent parity-even and parity-odd operators that can appear at this order, once redundancies due to identities of the Riemann and Ricci tensors, integration by parts, and field redefinitions are removed. We have shown that the resulting operator basis consists of five independent even-parity and three independent odd-parity structures, in complete agreement with the counting derived from on-shell scattering amplitudes.

The amplitude analysis not only confirms the operator counting but also provides a complementary and more physical viewpoint: local contact terms in the amplitude expansion directly encode the basis of higher-derivative operators in the EFT. The matching between these two approaches ensures that no additional hidden redundancies remain and that the six-derivative scalar-tensor EFT presented here is both minimal and complete.

The inclusion of six-derivative operators provides a framework for systematically studying subleading curvature corrections to scalar--tensor interactions, which may arise from integrating out heavy degrees of freedom in UV-complete theories such as string theory. The parity-odd invariants identified here also furnish a controlled setting in which to explore parity-violating effects in gravitational phenomena --- potentially relevant for signatures such as gravitational wave birefringence or chiral instabilities in the early universe.

It is natural to extend the present analysis in several directions. First, enumerating the independent operators at eight-derivative order would expose the next layer of higher-curvature corrections; this classification has already been carried out for shift-symmetric scalar theories in \cite{Ruhdorfer:2019qmk}. Second, the explicit six-derivative action derived here can serve as the starting point for investigating the perturbative stability and phenomenological implications of higher-derivative corrections in cosmological and black-hole backgrounds.

However, perhaps the most interesting new direction is to systematically explore the form of matter couplings order by order in the derivative expansion. Recall that the on-shell actions derived in this paper only describe the most general scalar–tensor dynamics away from external sources. As discussed in the text, couplings to matter are not preserved under the field redefinitions used to cast the effective action into its on-shell form at a given derivative order. One might regard matter couplings as irrelevant in contexts such as cold inflation — where matter is diluted — or for gravitational waves propagating through vacuum. But this is too hasty. Reheating physics is, of course, sensitive to how the inflaton couples to matter, and gravitational wave spectra may depend on source frequencies, and therefore on the detailed form of the coupling. We have also seen that nontrivial interactions between the scalar field and electromagnetism can complicate the interpretation of multi-messenger constraints on gravity from neutron-star events \cite{Mironov:2024idn,Babichev:2024kfo}.

The discussion above applies to standard Wilsonian EFTs, in which the degrees of freedom integrated out are heavy but, in principle, accessible and the resulting effective action is local. When the effective couplings to matter instead arise from genuinely inaccessible sectors or from complicated many-body or environmental processes, it is more appropriate to describe them using open system tools. In such open EFTs the  interactions are encoded in non-local response and noise kernels, and the local field redefinitions  used to cast a closed EFT into an on-shell basis are no longer generically available, but this discussion is beyond the scope of this work.

\section*{Acknowledgments} We would like to thank David Stefanyszyn and Philipp Neckam for useful discussions.  The work of EB was supported by ANR grant StronG (No. ANR-22-CE31-0015-01). SB was supported by the Austrian Science Fund, FWF, project no. P34562. MM is supported Kavli IPMU which was established by
the World Premier International Research Center Initiative (WPI), MEXT, Japan. MM is also grateful for the hospitality of Perimeter
Institute where part of this work was carried out. Her visit to Perimeter Institute was supported by a
grant from the Simons Foundation (1034867, Dittrich). Research at Perimeter Institute is supported in part by the Government of Canada through the Department of Innovation, Science and Economic Development and by the Province of Ontario through the Ministry of Colleges and Universities. AP is supported by STFC consolidated grant number ST/T000732/1. For the purpose of open access, the authors have applied a CC BY public copyright license to any Author Accepted Manuscript version arising. No new data were created during this study.

\appendix

\section{Identities}
		\label{identities}

\noindent\textbf{Symmetries and Identities of the Riemann Tensor}\vspace{-0.7em}
\begin{align}
    & \text{Pairwise exchange symmetry}\!: R_{\a\b\g\d}= R_{\g\d\a\b} \\
    & \text{Antisymmetry in the first two and last two indices}\!: R_{\a\b\g\d}= -R_{\b\a\g\d}=-R_{\a\b\d\g} \\
    & \text{Cyclicity}\footnotemark \!: R_{\a\b\g\d}+ R_{\a\d\b\g}+R_{\a\g\d\b}=0 \\
    & \text{Differential Bianchi identity}\!: \nabla_\zeta R_{\a\b\g\d}+\nabla_\d R_{\a\b\zeta\g}+\nabla_\g R_{\a\b\d\zeta}=0 
\end{align}
\footnotetext{aka algebraic Bianchi identity}
\noindent\textbf{Symmetries and Identities of the Ricci Tensor}\vspace{-0.7em}
\begin{align}
    & \text{Symmetry}\!: R_{\a\b}= R_{\b\a} \\
    & \text{Contracted differential Bianchi identity}\!: \nabla_\g R_{\a\b}-\nabla_\b R_{\a\g}+\nabla_\d R^\d{}_{\a\b\g}=0  \\
    & \text{Twice-contracted differential Bianchi identity}\!: \nabla^\a R_{\a\b}-\tfrac12 \nabla_\b R=0 
\end{align}

\noindent\textbf{Ricci Identity for}\vspace{-0.7em}
\begin{align}\label{commut_covd}
    &\text{a scalar:}\: [\na_\a,\na_\b]\phi=-\,T^\g_{\a\b}\,\pa_\g\phi=0 \ \ \because\, \text{the torsion tensor}\ T^\g_{\a\b}=0 \nn\\
    &\text{a vector:}\: [\na_\a,\na_\b]V_\g=R^\d{}_{\g\a\b}\,V_\d \nn\\
    &\text{a rank-2 tensor:}\: [\na_\a,\na_\b]A^\g{}_\z=R^\d{}_{\z\a\b}\,A^\g{}_\d -R^\g{}_{\d\a\b}\,A^\d{}_\z 
\end{align}
\vspace{-0.9em}

\noindent\textbf{Dimensionally Dependent Identities (DDIs)}\vspace{0.5em}

\noindent\textbf{Even Parity DDIs}\vspace{-0.5em}
\allowdisplaybreaks
\begin{small}
\begin{align}
&R_{\alpha\gamma\delta\varepsilon}
R_{\beta}{}^{\gamma\delta\varepsilon}
= 
2\,R_{\alpha\gamma} R_{\beta}{}^{\gamma}
- g_{\alpha\beta} R_{\gamma\delta} R^{\gamma\delta}
- R_{\alpha\beta} R
+ \frac{1}{4}\, g_{\alpha\beta} R^{2} 
+ 2\, R_{\gamma\delta}
  R_{\alpha}{}^{\gamma}{}_{\beta}{}^{\delta}
+ \frac{1}{4}\, g_{\alpha\beta}
  R_{\gamma\delta\varepsilon\zeta}
  \label{DDIeven1}
  R^{\gamma\delta\varepsilon\zeta} \\
&R^{\alpha}{}_{\varepsilon}{}^{\gamma}{}_{\zeta}
R_{\alpha\beta\gamma\delta}
R^{\beta\varepsilon\delta\zeta}=
-4\,R^{\alpha}{}_{\gamma}
R_{\alpha\beta}
R^{\beta\gamma}
+\frac{9}{2}\,
R^{\alpha\beta}
R_{\alpha\beta}R
-
\frac{5}{8}\,R^{3}
-3\,
R_{\alpha\beta}
R_{\gamma\delta}
R^{\alpha\gamma\beta\delta}
-\frac{3}{8}\,R
R^{\alpha\beta\gamma\delta}
R_{\alpha\beta\gamma\delta}
\nn\\ &\qquad\qquad\qquad\qquad\qquad+\frac{1}{2}\,
R^{\alpha\beta}{}_{\varepsilon\zeta}
R_{\alpha\beta\gamma\delta}
R^{\gamma\delta\varepsilon\zeta}
  \label{DDIeven2}
\end{align}
\end{small}
We note that DDIs may be derived in either a Riemann or Weyl tensor basis. We show equations ~(\ref{DDIgeneral})-(\ref{dualDDI3C}) below, to give one possible origin of the DDIs -- in terms of Weyl tensor relations.

One may start with the relations
\begin{equation}
\label{DDIgeneral}
T^{[\gamma\delta}{}_{[\mu\nu} \delta_{\beta]}^{\alpha]}=0,
\end{equation}
which are valid in 4D for any tensor $T^{\gamma\delta}{}_{\epsilon\eta}$ which is antisymmetric on upper and lower indices, see ref.~\cite{Edgar:2001vv}.
The square brackets used above imply antisymmetrisation over indices, 
\begin{equation*}
T_{\left[\alpha_1 \ldots \alpha_p\right]}=\frac{1}{p!} \delta_{\alpha_1 \ldots \alpha_p}^{\beta_1 \ldots \beta_p} T_{\beta_1 \ldots \beta_p}.
\end{equation*}
Since the Weyl tensor $C^{\gamma\delta}{}_{\mu\nu}$ satisfies the desired symmetries, on substituting $T^{\gamma\delta}{}_{\mu\nu}$ in~(\ref{DDIgeneral}) with it and contracting with $C^{\mu\nu}{}_{\gamma\delta}$ one obtains,
\begin{equation}
\label{DDI2C}
C_\alpha{ }^{\gamma \delta \epsilon} C_{\beta \gamma \delta \epsilon}-\frac{1}{4} g_{\alpha \beta} C_{\gamma \delta \epsilon \eta}^2=0.
\end{equation}
This expression may be rewritten in the form~(\ref{DDIeven1}) by expressing the Weyl tensor in terms of the Riemann tensor.

Similarly, on replacing $T^{\gamma\delta}{}_{\mu\nu}$ in ~(\ref{DDIgeneral}) with the Weyl tensor $C^{\gamma\delta}{}_{\mu\nu}$ and contracting it with $C^{\mu\nu}{}_{\lambda\alpha} C^{\beta\lambda}{}_{\gamma\delta}$, one obtains 
\begin{equation}
C^{\alpha \beta \gamma \delta} C_\alpha{}^\epsilon{}_\gamma{}^\eta C_{\beta \eta \delta \epsilon} - \frac{1}{4} C_{\alpha \beta}{}^{\epsilon \eta} C^{\alpha \beta \gamma \delta} C_{\gamma \delta \epsilon \eta} =0.
\end{equation}
Expressing the Weyl tensors here in terms of the Riemann tensor and combining with ~(\ref{DDI2C}) provides one route to obtaining~(\ref{DDIeven2})
\vspace{0.5em}

\noindent\textbf{Odd Parity DDIs}
\\~
The number of independent identities in the odd parity case is larger as compared to the even-parity case due to the presence of the Levi-Civita tensor. Indeed, let us start with identities with one curvature tensor. 
First of all, the following identity is true, see, e.g.~\cite{Stephani:2003tm}, 
\begin{equation}
\label{ruse-lanczos}
	{}^*C_{\alpha\beta\gamma\delta} = C^*_{\alpha\beta\gamma\delta} \;\;  \Leftrightarrow \;\;
	{}^*C_{\gamma\delta\alpha\beta} =  {}^*C_{\alpha\beta\gamma\delta} \;\;  \Leftrightarrow \;\; 
	\epsilon_{\alpha\beta\mu\nu}C^{\mu\nu}{}_{\gamma\delta} = \epsilon_{\gamma\delta\mu\nu}C^{\mu\nu}{}_{\alpha\beta}
\end{equation}
where $^*C_{\alpha\beta\gamma\delta}$ and  $C^*_{\alpha\beta\gamma\delta}$ are the left dual and right dual of the Weyl tensor:
\begin{equation*}
 ^*C_{\alpha \beta \gamma \delta}=\frac{1}{2} \epsilon_{\alpha \beta \mu \nu} C_{\gamma \delta}^{\mu \nu}, \quad
 C_{\alpha \beta \gamma \delta}^*=\frac{1}{2} C_{\alpha \beta}^{\mu \nu} \epsilon_{\mu \nu \gamma \delta}.
\end{equation*}
Then, we can construct the following DDI involving only the Ricci tensor and the Ricci scalar,
\begin{equation}
\label{dualDDIR}
g^{\mu \nu} \epsilon_{\alpha[\beta \gamma \delta} R_{\mu \nu]}=0  \;\;  \Leftrightarrow \;\; 
3 \epsilon_{\alpha[\beta \gamma}{ }^\mu R_{\delta ] \mu}=\epsilon_{\beta\gamma \delta}{}^\mu R_{\alpha \mu}+\epsilon_{\alpha \beta \gamma\delta}  R.
\end{equation}
The algebraic Bianchi identity for the dual Weyl tensor is also satisfied. 
It follows from the identity 
$\epsilon_{\alpha[\beta \mu \nu} C_{\gamma \delta] \eta \varepsilon}=0$ with the indices $\mu$ and $\varepsilon$, and $\nu$ and $\eta$, contracted,
\begin{equation}
\label{dualBianchi}
\epsilon_{\mu \nu \alpha[\beta} C_{\gamma \delta]}{ }^{\mu \nu}=0 \;\; \Leftrightarrow \;\; {}^* C_{\alpha[\beta \gamma \delta]}=0.
\end{equation}
Next, we look at identities quadratic in the curvature tensor, similar to the even-parity case. 
An analogue of~(\ref{DDI2C}) can be found from~(\ref{DDIgeneral}), where in place of  $T^{\gamma\delta}{}_{\epsilon\eta}$ we write the Weyl tensor and we contract the obtained expression with ${}^*C^{\mu\nu}{}_{\gamma\delta}$. Then using~(\ref{ruse-lanczos}) we get,
\begin{equation}
\label{dualDDI2C}
{ }^* C_{\alpha \gamma\delta\epsilon} C^{\beta \gamma\delta\epsilon}=\frac{1}{4} \delta_\alpha^{\beta}\; {}^*C_{\gamma\delta\epsilon\eta} C^{\gamma\delta\epsilon\eta}.
\end{equation}
Finally, on replacing $T^{\gamma\delta}{}_{\mu\nu}$ in~(\ref{DDIgeneral}) with the Weyl tensor $C^{\gamma\delta}{}_{\mu\nu}$ and contracting with ${}^*C^{\mu\nu}{}_{\lambda\alpha} C^{\beta\lambda}{}_{\gamma\delta}$, we get,
\begin{equation}
\label{dualDDI3C}
C^{\alpha \beta \gamma \delta}\; C_\alpha{}^\epsilon{}_\gamma{}^\eta \;{}^*C_{\delta \epsilon \beta \eta} 
-\frac{1}{4} C_{\alpha \beta}{}^{\epsilon \eta}\; C^{\alpha \beta \gamma \delta} \;{}^*C_{\epsilon \eta \gamma \delta} =0.
\end{equation}
The above five independent identities (\ref{ruse-lanczos}) -- (\ref{dualDDI3C}), along with the symmetries of the Riemann and Weyl tensors, provide useful context. For our analysis, we make use of the following identities:
\allowdisplaybreaks
\begin{small}
\begin{align}
&\epsilon^{\delta\varepsilon\zeta\eta}
R_{\alpha\gamma\delta\varepsilon}
R_{\beta}{}^{\gamma}{}_{\zeta\eta}=
2\,\epsilon_{\alpha}{}^{\delta\varepsilon\zeta}
R_{\gamma\delta}
R_{\beta}{}^{\gamma}{}_{\varepsilon\zeta}
+\frac{1}{4}\,
\epsilon^{\varepsilon\zeta\eta\theta}
g_{\alpha\beta}
R^{\gamma\delta}{}_{\eta\theta}
R_{\gamma\delta\varepsilon\zeta}\\
&\epsilon_{\beta}{}^{\gamma\zeta\eta}
R_{\alpha\gamma\delta\varepsilon}
R^{\delta\varepsilon}{}_{\zeta\eta}=
\frac{1}{4}\,
\epsilon^{\varepsilon\zeta\eta\theta}
g_{\alpha\beta}
R^{\gamma\delta}{}_{\eta\theta}
R_{\gamma\delta\varepsilon\zeta}\\
&\epsilon^{\beta\gamma\delta\varepsilon}
\nabla_{\varepsilon}\!\left(
  \nabla_{\delta}\!\left(
    \nabla_{\gamma} R_{\alpha\beta}
  \right)
\right)=
-\frac{1}{2}\,
\epsilon_{\alpha}{}^{\gamma\varepsilon\zeta}
R^{\beta\delta}{}_{\varepsilon\zeta}
\nabla_{\delta} R_{\beta\gamma}
-
\epsilon_{\alpha}{}^{\gamma\delta\varepsilon}
R_{\beta\gamma}
\nabla_{\varepsilon} R^{\beta}{}_{\delta} \label{5DerOddDDI1}\\
&\epsilon^{\gamma\delta\varepsilon\zeta}
R_{\alpha}{}^{\beta}{}_{\varepsilon\zeta}
\nabla_{\delta} R_{\beta\gamma}=
\epsilon_{\alpha}{}^{\gamma\varepsilon\zeta}
R^{\beta\delta}{}_{\varepsilon\zeta}
\nabla_{\delta} R_{\beta\gamma}+2\,
\epsilon_{\alpha}{}^{\gamma\delta\varepsilon}
R_{\beta\gamma}
\nabla_{\varepsilon} R^{\beta}{}_{\delta} \label{5DerOddDDI2}\\
&\epsilon^{\delta\varepsilon\zeta\eta}
R_{\beta\gamma\delta\varepsilon}
\nabla_{\alpha}
R^{\beta\gamma}{}_{\zeta\eta}=
2\,\epsilon_{\alpha}{}^{\varepsilon\zeta\eta}
R_{\beta\gamma\delta\varepsilon}
\nabla^{\delta}
R^{\beta\gamma}{}_{\zeta\eta}
-4\,\epsilon_{\alpha}{}^{\gamma\varepsilon\zeta}
R^{\beta\delta}{}_{\varepsilon\zeta}
\nabla_{\delta} R_{\beta\gamma} \label{5DerOddDDI3} \qquad\qquad\qquad\quad\\
&\epsilon^{\delta\zeta\eta\theta}
R^{\alpha}{}_{\varepsilon}{}^{\gamma}{}_{\zeta}
R_{\alpha\beta\gamma\delta}
R^{\beta\varepsilon}{}_{\eta\theta}=
-
\epsilon^{\beta\delta\varepsilon\zeta}
R_{\alpha\beta}
R_{\gamma\delta}
R^{\alpha\gamma}{}_{\varepsilon\zeta}
-\frac{3}{8}\,
\epsilon^{\gamma\delta\varepsilon\zeta}
R
R^{\alpha\beta}{}_{\varepsilon\zeta}
R_{\alpha\beta\gamma\delta}
+\frac{1}{2}\,
\epsilon^{\varepsilon\zeta\eta\theta}
R^{\alpha\beta}{}_{\varepsilon\zeta}
R_{\alpha\beta\gamma\delta}
R^{\gamma\delta}{}_{\eta\theta} \\
&\epsilon^{\delta\varepsilon\eta\theta}
R^{\alpha}{}_{\varepsilon}{}^{\gamma}{}_{\zeta}
R_{\alpha\beta\gamma\delta}
R^{\beta\zeta}{}_{\eta\theta}=
-
\epsilon^{\beta\delta\varepsilon\zeta}
R_{\alpha\beta}
R_{\gamma\delta}
R^{\alpha\gamma}{}_{\varepsilon\zeta}
-\frac{1}{4}\,
\epsilon^{\gamma\delta\varepsilon\zeta}
R
R^{\alpha\beta}{}_{\varepsilon\zeta}
R_{\alpha\beta\gamma\delta}
+\frac{1}{4}\,
\epsilon^{\varepsilon\zeta\eta\theta}
R^{\alpha\beta}{}_{\varepsilon\zeta}
R_{\alpha\beta\gamma\delta}
R^{\gamma\delta}{}_{\eta\theta}\\
&\epsilon^{\delta\zeta\eta\theta}
R^{\alpha\beta}{}_{\varepsilon\zeta}
R_{\alpha\beta\gamma\delta}
R^{\gamma\varepsilon}{}_{\eta\theta}=
-\frac{1}{4}\,
\epsilon^{\gamma\delta\varepsilon\zeta}
R
R^{\alpha\beta}{}_{\varepsilon\zeta}
R_{\alpha\beta\gamma\delta}
+\frac{1}{2}\,
\epsilon^{\varepsilon\zeta\eta\theta}
R^{\alpha\beta}{}_{\varepsilon\zeta}
R_{\alpha\beta\gamma\delta}
R^{\gamma\delta}{}_{\eta\theta}\\
&\epsilon_{\gamma\delta}{}^{\varepsilon\zeta}
R_{\alpha\beta\varepsilon\zeta}=
2\,
\epsilon_{\alpha\gamma\delta}{}^{\eta}
R_{\beta\eta}
-\epsilon_{\alpha\delta}{}^{\eta\theta}
R_{\beta\gamma\eta\theta}
+\epsilon_{\alpha\gamma}{}^{\eta\theta}
R_{\beta\delta\eta\theta}\\
&\epsilon_{\alpha\beta}{}^{\delta\varepsilon}
\nabla_{\varepsilon} R_{\gamma\delta}=
\frac{1}{2}\,
\epsilon_{\alpha\beta\gamma}{}^{\zeta}
\nabla_{\zeta} R
-\epsilon_{\beta\gamma}{}^{\zeta\eta}
\nabla_{\eta} R_{\alpha\zeta}
+\epsilon_{\alpha\gamma}{}^{\zeta\eta}
\nabla_{\eta} R_{\beta\zeta}\\
&\epsilon_{\beta}{}^{\delta\varepsilon\zeta}
R_{\gamma\delta}
R_{\alpha}{}^{\gamma}{}_{\varepsilon\zeta}=
\epsilon_{\alpha}{}^{\delta\varepsilon\zeta}
R_{\gamma\delta}
R_{\beta}{}^{\gamma}{}_{\varepsilon\zeta}
\end{align}
\end{small}

		\section{Off-shell Six-Derivative Scalar-Tensor EFT}
		\label{offshell_app}

On generating all possible six-derivative terms of even parity independent up to integration by parts, as explained in sec. \ref{even_onshell}, we get the following eighty-nine terms, where the coefficients $e_n(\phi)=\tilde e_n(\phi)/\Lambda^2$ with $\tilde e_n(\phi)$ being arbitrary algebraic scalar functions of $\phi$ and $\Lambda$ a scaling parameter with mass dimension 1:\footnote{Appendix A.2 of \cite{Aguilar-Gutierrez:2023kfn} presents a basis of six-derivative pure-gravity terms, but it contains redundancies and omissions. Seventeen six-derivative invariants are listed, whereas only fifteen are independent. Their redundant terms include $R^{(1)}_6=R^{(2)}_6+R^{(6)}_6$, $R^{(15)}_6= 2R^{(12)}_6-2R^{(11)}_6+\tfrac14 R^{(10)}_6+2R^{(13)}_6+\tfrac14 R^{(14)}_6$, $R^{(17)}_6= -4R^{(12)}_6+\tfrac92 R^{(11)}_6-\tfrac58 R^{(10)}_6-3R^{(13)}_6-\tfrac38 R^{(14)}_6+\tfrac12 R^{(16)}_6$, while $R^{(5)}_6=0$, and $R^{(3)}_6$ is actually an eight-derivative term. Conversely, three independent terms missing from their list are $R_{\alpha\beta\gamma\delta} \nabla^\delta \nabla^\beta R^{\alpha\gamma}$, $R_{\alpha\beta} \nabla^\beta \nabla^\alpha R$ and $\Box^2 R$.}
\allowdisplaybreaks
\begin{small}
\begin{align}\label{89_even}
\allowdisplaybreaks
&e_1(\phi)R_{\alpha\beta\gamma\delta} R^{\alpha\beta}{}_{\zeta\eta} R^{\gamma\delta\zeta\eta}
,&&\!\!\!\!e_2(\phi)\nabla_\zeta R_{\alpha\beta\gamma\delta} \, \nabla^\zeta R^{\alpha\beta\gamma\delta}
,&&\!\!\!\!e_3(\phi) R_{\alpha\beta\gamma\delta} R^{\alpha\beta\gamma\delta}R
,\nn\\
&e_4(\phi) R_{\alpha \gamma \beta \delta}R^{\alpha \beta} R^{\gamma \delta}
,&&\!\!\!\!e_5(\phi)R_{\alpha\beta\gamma\delta} \nabla^\delta \nabla^\beta R^{\alpha\gamma}
,&&\!\!\!\!e_6(\phi)R_{\alpha \beta} R^\beta{}_\gamma R^{\alpha\gamma}, \nn\\
&e_7(\phi)\nabla_\gamma R_{\alpha\beta} \, \nabla^\gamma R^{\alpha\beta},&&\!\!\!\!e_8(\phi) \nabla_\gamma R_{\alpha\beta}\,\nabla^\beta R^{\alpha\gamma},&&\!\!\!\!e_9(\phi)R_{\alpha\beta}\, \bo R^{\alpha\beta}
,\nn\\
&e_{10}(\phi)R_{\alpha\beta} R^{\alpha\beta} R
,&&\!\!\!\!e_{11}(\phi)R_{\alpha\beta} \nabla^\beta \nabla^\alpha R
,&&\!\!\!\!e_{12}(\phi)R^3
,\nn\\
&e_{13}(\phi)\nabla_\alpha R \, \nabla^\alpha R
,&&\!\!\!\!e_{14}(\phi)R \, \bo R
,&&\!\!\!\!e_{15}(\phi)\,\bo^2 R,\nn\\[-3mm]
 &\quad\,\mathclap{\rule{1cm}{0.4pt}}\nn\\[-2mm]
&e_{16}(\phi)R_{\alpha\beta\gamma\delta} R^{\alpha\beta\gamma\delta} \nabla_{\zeta}\phi \nabla^{\zeta}\phi,&&\!\!\!\!e_{17}(\phi)R_{\alpha\beta\gamma\delta} R^{\alpha\beta\gamma\delta}\, \bo\phi
,&&\!\!\!\!e_{18}(\phi)R^{\alpha\beta\gamma\delta} \nabla_{\zeta} R_{\alpha\beta\gamma\delta} \nabla^{\zeta}\phi
, \nn\\
&e_{19}(\phi)R_{\alpha\beta\gamma\delta} R^{\beta\delta} \nabla^{\alpha}\phi \nabla^{\gamma}\phi
,&&\!\!\!\!e_{20}(\phi)R_{\alpha\beta\gamma\delta} R^{\alpha\gamma} \nabla^{\delta}\nabla^{\beta}\phi
,&&\!\!\!\!e_{21}(\phi)R_{\alpha\beta\gamma\delta}\nabla^{\delta} R^{\beta\gamma} \nabla^{\alpha}\phi 
, \nn\\
&e_{22}(\phi)R_{\alpha\beta\gamma\delta} \nabla^{\alpha}\phi \nabla^{\gamma}\phi \nabla^{\delta}\nabla^{\beta}\phi,&&\!\!\!\!e_{23}(\phi)R_{\alpha\beta\gamma\delta} \nabla^{\gamma}\nabla^{\alpha}\phi \nabla^{\delta}\nabla^{\beta}\phi,&&\!\!\!\!e_{24}(\phi)R_{\alpha\beta} R^{\alpha\beta} \nabla_{\gamma}\phi \nabla^{\gamma}\phi,\nn\\
&e_{25}(\phi)R_{\alpha\beta} R^{\alpha}{}_{\gamma} \nabla^{\beta}\phi\nabla^{\gamma}\phi ,&&\!\!\!\!e_{26}(\phi)R_{\alpha\beta} \nabla_{\gamma} R^{\alpha\beta} \nabla^{\gamma}\phi,&&\!\!\!\!e_{27}(\phi)
R_{\alpha\beta} \nabla^{\beta} R^\alpha{}_\gamma\nabla^{\gamma}\phi ,\nn\\
&e_{28}(\phi)R_{\alpha\beta}R^\alpha{}_\gamma  \nabla^{\gamma} \nabla^{\beta}\phi,&&\!\!\!\!e_{29}(\phi)R_{\alpha\beta} R \, \nabla^{\alpha}\phi \, \nabla^{\beta}\phi,&&\!\!\!\!e_{30}(\phi)R_{\alpha\beta} \, \nabla^{\alpha} R \, \nabla^{\beta}\phi,\nn\\
&e_{31}(\phi)R_{\alpha\beta} R^{\alpha\beta} \, \bo\phi,&&\!\!\!\!e_{32}(\phi)
R_{\alpha\beta} R \, \nabla^{\beta} \nabla^{\alpha} \phi,&&\!\!\!\!e_{33}(\phi)R_{\alpha\beta} \, \nabla^{\alpha}\phi \, \nabla^{\beta}\phi\, \nabla_{\gamma}\phi \, \nabla^{\gamma}\phi,\nn\\
&e_{34}(\phi)
R_{\alpha\beta} \, \nabla^{\alpha} \nabla^{\beta}\phi\, \nabla_{\gamma}\phi \, \nabla^{\gamma}\phi,&&\!\!\!\!e_{35}(\phi)
R_{\alpha\beta} \, \nabla^{\alpha}\phi \, \nabla^{\beta}\phi \, \bo \phi,&&\!\!\!\!e_{36}(\phi)
R_{\alpha\beta} \, \nabla^{\alpha}\phi\, \nabla^{\beta} \nabla_{\gamma} \phi \, \nabla^{\gamma}\phi , \nn\\
&e_{37}(\phi)\nabla_{\gamma} R_{\alpha\beta} \,\nabla^{\alpha}\phi \, \nabla^{\beta}\phi \,  \nabla^{\gamma}\phi,&&\!\!\!\!e_{38}(\phi)
R_{\alpha\beta} \, \nabla^{\beta}\nabla^{\alpha}\phi \, \bo\phi,&&\!\!\!\!e_{39}(\phi)
R_{\alpha\beta} \, \nabla_{\gamma}\nabla^{\alpha}\phi \, \nabla^{\gamma}\nabla^{\beta}\phi,\nn\\
&e_{40}(\phi)
\nabla_{\alpha} R_{\beta\gamma} \, \nabla^{\alpha}\phi \, \nabla^{\gamma}\nabla^{\beta}\phi,&&\!\!\!\!e_{41}(\phi)\nabla_{\gamma} R_{\alpha\beta} \,\nabla^{\alpha}\phi \,  \nabla^{\gamma} \nabla^{\beta}\phi,&&\!\!\!\!e_{42}(\phi)
R_{\alpha\beta} \, \nabla^{\alpha}\phi \, \nabla^{\beta}\, \bo\phi,\nn\\
&e_{43}(\phi)
R_{\alpha\beta} \, \nabla_{\gamma}\phi \, \nabla^{\alpha} \nabla^{\beta} \nabla^{\gamma}\phi,&&\!\!\!\!e_{44}(\phi)\,\bo R_{\alpha\beta}\,
\nabla^{\alpha}\phi \, \nabla^{\beta}\phi,&&\!\!\!\!e_{45}(\phi)\nabla_{\gamma} R_{\alpha\beta}\,\nabla^{\gamma} \nabla^{\beta} \nabla^{\alpha} \phi \, ,\nn\\
&e_{46}(\phi)
R_{\alpha\beta} \, \nabla^{\beta} \nabla^{\alpha} \bo \phi,&&\!\!\!\!e_{47}(\phi)\,\bo R_{\alpha\beta}
\nabla^{\alpha} \nabla^{\beta} \phi ,&&\!\!\!\!e_{48}(\phi)
R^2 \, \nabla_{\alpha} \phi \, \nabla^{\alpha} \phi,\nn\\
&e_{49}(\phi)R \,\nabla_{\alpha} R\, \nabla^{\alpha}\phi,&&\!\!\!\!e_{50}(\phi)
R^2 \, \bo \phi,&&\!\!\!\!e_{51}(\phi)
R \, (\nabla_{\alpha}\phi \, \nabla^{\alpha}\phi)^2,\nn\\
&e_{52}(\phi)
R\, \nabla_{\alpha}\phi \, \nabla^{\alpha}\phi \, \bo \phi,&&\!\!\!\!e_{53}(\phi)R \, \nabla_{\alpha}\phi\, \nabla_{\beta}\phi \, \nabla^{\alpha} \nabla^{\beta}\phi ,&&\!\!\!\!e_{54}(\phi)\nabla_{\alpha} R \,
\nabla^{\alpha}\phi \,  \nabla_{\beta}\phi \, \nabla^{\beta}\phi,\nn\\
&e_{55}(\phi)
R \, (\bo \phi)^2 ,&&\!\!\!\!e_{56}(\phi)
R \, \nabla_{\beta} \nabla_{\alpha}\phi \, \nabla^{\beta} \nabla^{\alpha} \phi,&&\!\!\!\!e_{57}(\phi)R \, \nabla_{\alpha}\phi\,\nabla^{\alpha} \bo \phi ,\nn\\
&e_{58}(\phi) \nabla_{\alpha} R \,
\nabla^{\alpha}\phi \, \bo \phi,&&\!\!\!\!e_{59}(\phi)
\nabla_{\alpha} R \,\nabla_{\beta}\phi\,\nabla^{\beta} \nabla^{\alpha} \phi,&&\!\!\!\!e_{60}(\phi)\,\bo R\,\nabla_{\alpha}\phi \, \nabla^{\alpha}\phi \, ,\nn\\
&e_{61}(\phi)\nabla_{\beta} \nabla_{\alpha} R \, \nabla^{\alpha}\phi \, \nabla^{\beta}\phi,&&\!\!\!\!e_{62}(\phi)\nabla^{\alpha}R\,
\nabla_{\alpha}\bo \phi,&&\!\!\!\!e_{63}(\phi)
R \, \bo^2 \phi,\nn\\
&e_{64}(\phi)\,
\bo R \, \bo \phi,&&\!\!\!\!e_{65}(\phi)\nabla_{\beta} \nabla_{\alpha} R\,\nabla^{\beta} \nabla^{\alpha} \phi ,&&\!\!\!\!e_{66}(\phi)
\nabla_{\alpha}\bo R \, \nabla^{\alpha} \phi,\nn\\[-3mm]
&\quad\,\mathclap{\rule{1cm}{0.4pt}}\nn\\[-2mm]
&e_{67}(\phi) (\nabla_\alpha \phi \, \nabla^\alpha \phi)^3 ,&&\!\!\!\!e_{68}(\phi) (\nabla_\alpha \phi \, \nabla^\alpha \phi)^2\, \bo \phi,&&\!\!\!\!e_{69}(\phi)\nabla_\alpha \phi \, \nabla^\alpha \phi \, \nabla_\beta \phi\,\nabla_\gamma\phi \, \nabla^\gamma \nabla^\beta \phi,\nn\\
&e_{70}(\phi)\nabla_\alpha \phi \, \nabla^\alpha \phi \, (\bo \phi)^2,&&\!\!\!\!e_{71}(\phi) \nabla_\alpha \phi \, \nabla^\alpha \phi \, \nabla_\gamma \nabla_\beta \phi \, \nabla^\gamma \nabla^\beta \phi,&&\!\!\!\!e_{72}(\phi)\nabla^\alpha \phi \, \nabla^\beta \phi \, \nabla_\gamma \nabla_\beta \phi \, \nabla^\gamma \nabla_\alpha \phi,\nn\\
&e_{73}(\phi)\nabla_\alpha \phi \,\nabla_\beta \phi \, \nabla^\beta \nabla^\alpha \phi \,  \bo \phi ,&&\!\!\!\!e_{74}(\phi)\nabla_\alpha \phi \, \nabla^\alpha \phi \,\nabla_\beta \phi \, \nabla^\beta \bo \phi \, ,&&\!\!\!\!e_{75}(\phi)\nabla_\alpha \phi \, \nabla_\beta \phi \, \nabla_\gamma \phi \,\nabla^\gamma \nabla^\beta \nabla^\alpha \phi , \nn\\
&e_{76}(\phi)(\bo\phi)^3,&&\!\!\!\!e_{77}(\phi)\nabla_{\alpha}\nabla_{\beta}\phi\,\nabla^{\alpha}\nabla^{\beta}\phi\,\bo\phi,&&\!\!\!\!e_{78}(\phi)\nabla_{\beta}\nabla_{\alpha}\phi\,\nabla_{\gamma}\nabla^{\beta}\phi\,\nabla^{\gamma}\nabla^{\alpha}\phi,\nn\\
&e_{79}(\phi)\nabla_\alpha \phi \, \bo \phi \,\nabla^\alpha \bo \phi ,&&\!\!\!\!e_{80}(\phi)\nabla_\alpha \phi \,\nabla_\beta \nabla^\alpha \phi\, \nabla^\beta \bo \phi ,&&\!\!\!\!e_{81}(\phi)\nabla_\alpha \phi \,\nabla_\gamma \nabla_\beta \phi\, \nabla^\gamma \nabla^\beta \nabla^\alpha \phi ,\nn\\
&e_{82}(\phi)\nabla_\alpha \phi \, \nabla^\alpha \phi \, \bo^2\phi,&&\!\!\!\!e_{83}(\phi)\nabla_\alpha \phi \,\nabla_\beta \phi \, \nabla^\beta \nabla^\alpha \bo\phi ,&&\!\!\!\!e_{84}(\phi)\nabla_{\a}\bo\phi\,\nabla^{\a}\bo\phi,\nn\\
&e_{85}(\phi)\nabla_{\gamma}\nabla_{\beta}\nabla_{\alpha}\phi\,\nabla^{\gamma}\nabla^{\beta}\nabla^{\alpha}\phi,&&\!\!\!\!e_{86}(\phi)\,\bo\phi\,\bo^2\phi,&&\!\!\!\!e_{87}(\phi)\nabla_{\beta}\nabla_{\alpha}\phi\,\nabla^{\beta}\nabla^{\alpha}\bo\phi,\nn\\
&e_{88}(\phi)\nabla_\alpha \phi\,\nabla^\alpha \bo^2 \phi ,&&\!\!\!\!e_{89}(\phi)\,\bo^3\phi.
\end{align}
\end{small}
In the above list, the first fifteen terms do not involve derivatives of the scalar field $\phi$; terms sixteen through sixty-six contain both curvature tensors/scalars and $\phi$; and terms sixty-seven through eighty-nine are purely scalar. Upon integrating by parts to remove redundant contributions, forty independent terms remain. Keeping those consisting of the d’Alembertian of $\phi$ and with the minimal derivative order per field, yields the following set of forty terms:
\allowdisplaybreaks
\begin{small}
\begin{align}\label{even_offshell}
\allowdisplaybreaks
&e_1(\phi)R_{\alpha\beta\gamma\delta} R^{\alpha\beta}{}_{\zeta\eta} R^{\gamma\delta\zeta\eta}
,&&\!\!\!\!e_3(\phi) R_{\alpha\beta\gamma\delta} R^{\alpha\beta\gamma\delta}R
,&&\!\!\!\!e_4(\phi) R_{\alpha \gamma \beta \delta}R^{\alpha \beta} R^{\gamma \delta}
,\nn\\
&e_5(\phi)R_{\alpha\beta\gamma\delta} \nabla^\delta \nabla^\beta R^{\alpha\gamma}
,&&\!\!\!\!e_6(\phi)R_{\alpha \beta} R^\beta{}_\gamma R^{\alpha\gamma} 
,&&\!\!\!\!e_7(\phi)\nabla_\gamma R_{\alpha\beta} \, \nabla^\gamma R^{\alpha\beta}
,\nn\\
&e_{10}(\phi)R_{\alpha\beta} R^{\alpha\beta} R
,&&\!\!\!\!e_{12}(\phi)R^3
,&&\!\!\!\!e_{13}(\phi)\nabla_\alpha R \, \nabla^\alpha R
,\nn\\[-3mm]
 &\quad\,\mathclap{\rule{1cm}{0.4pt}}\nn\\[-2mm]
&e_{16}(\phi)R_{\alpha\beta\gamma\delta} R^{\alpha\beta\gamma\delta} \nabla_{\zeta}\phi \nabla^{\zeta}\phi,&&\!\!\!\!e_{17}(\phi)R_{\alpha\beta\gamma\delta} R^{\alpha\beta\gamma\delta}\, \bo\phi
,&&\!\!\!\!e_{19}(\phi)R_{\alpha\beta\gamma\delta} R^{\beta\delta} \nabla^{\alpha}\phi \nabla^{\gamma}\phi
,\nn\\
&e_{22}(\phi)R_{\alpha\beta\gamma\delta} \nabla^{\alpha}\phi \nabla^{\gamma}\phi \nabla^{\delta}\nabla^{\beta}\phi,&&\!\!\!\!e_{24}(\phi)R_{\alpha\beta} R^{\alpha\beta} \nabla_{\gamma}\phi \nabla^{\gamma}\phi,&&\!\!\!\!e_{25}(\phi)R_{\alpha\beta} R^{\alpha}{}_{\gamma} \nabla^{\beta}\phi\nabla^{\gamma}\phi ,\nn\\
&e_{27}(\phi)
R_{\alpha\beta} \nabla^{\beta} R^\alpha{}_\gamma\nabla^{\gamma}\phi ,&&\!\!\!\!e_{29}(\phi)R_{\alpha\beta} R \, \nabla^{\alpha}\phi \, \nabla^{\beta}\phi,&&\!\!\!\!e_{30}(\phi)R_{\alpha\beta} \, \nabla^{\alpha} R \, \nabla^{\beta}\phi,\nn\\
&e_{31}(\phi)R_{\alpha\beta} R^{\alpha\beta} \, \bo\phi,&&\!\!\!\!e_{33}(\phi)R_{\alpha\beta} \, \nabla^{\alpha}\phi \, \nabla^{\beta}\phi\, \nabla_{\gamma}\phi \, \nabla^{\gamma}\phi,&&\!\!\!\!e_{34}(\phi)
R_{\alpha\beta} \, \nabla^{\alpha} \nabla^{\beta}\phi\, \nabla_{\gamma}\phi \, \nabla^{\gamma}\phi,\nn\\
&e_{35}(\phi)
R_{\alpha\beta} \, \nabla^{\alpha}\phi \, \nabla^{\beta}\phi \, \bo \phi,&&\!\!\!\!e_{38}(\phi)
R_{\alpha\beta} \, \nabla^{\beta}\nabla^{\alpha}\phi \, \bo\phi,&&\!\!\!\!e_{40}(\phi)
\nabla_{\alpha} R_{\beta\gamma} \, \nabla^{\alpha}\phi \, \nabla^{\gamma}\nabla^{\beta}\phi,\nn\\
&e_{41}(\phi)\nabla_{\gamma} R_{\alpha\beta} \,\nabla^{\alpha}\phi \,  \nabla^{\gamma} \nabla^{\beta}\phi,&&\!\!\!\!e_{48}(\phi)
R^2 \, \nabla_{\alpha} \phi \, \nabla^{\alpha} \phi,&&\!\!\!\!e_{50}(\phi)
R^2 \, \bo \phi,\nn\\
&e_{51}(\phi)
R \, (\nabla_{\alpha}\phi \, \nabla^{\alpha}\phi)^2,&&\!\!\!\!e_{52}(\phi)
R\, \nabla_{\alpha}\phi \, \nabla^{\alpha}\phi \, \bo \phi,&&\!\!\!\!e_{55}(\phi)
R \, (\bo \phi)^2 ,\nn\\
&e_{57}(\phi)R \, \nabla_{\alpha}\phi\,\nabla^{\alpha} \bo \phi ,&&\!\!\!\!e_{60}(\phi)\,\bo R\,\nabla_{\alpha}\phi \, \nabla^{\alpha}\phi ,&&\!\!\!\!e_{66}(\phi)
\nabla_{\alpha}\bo R \, \nabla^{\alpha} \phi,\nn\\[-3mm]
&\quad\,\mathclap{\rule{1cm}{0.4pt}}\nn\\[-2mm]
&e_{67}(\phi) (\nabla_\alpha \phi \, \nabla^\alpha \phi)^3 ,&&\!\!\!\!e_{68}(\phi) (\nabla_\alpha \phi \, \nabla^\alpha \phi)^2\, \bo \phi,&&\!\!\!\!e_{70}(\phi)\nabla_\alpha \phi \, \nabla^\alpha \phi \, (\bo \phi)^2,\nn\\
&e_{71}(\phi) \nabla_\alpha \phi \, \nabla^\alpha \phi \, \nabla_\gamma \nabla_\beta \phi \, \nabla^\gamma \nabla^\beta \phi,&&\!\!\!\!e_{74}(\phi)\nabla_\alpha \phi \, \nabla^\alpha \phi \,\nabla_\beta \phi \, \nabla^\beta \bo \phi,&&\!\!\!\!e_{76}(\phi)(\bo\phi)^3,\nn\\
&e_{84}(\phi)\nabla_{\a}\bo\phi\,\nabla^{\a}\bo\phi.
\end{align}
\end{small}
On constructing all six-derivative odd-parity terms independent before integration by parts, as outlined in Sec.~\ref{odd_onshell}, we obtain the following sixteen terms, with the coefficient functions $o_n(\phi)=\tilde o_n(\phi)/\Lambda^2$ where $\tilde o_n(\phi)$ are general algebraic scalar functions of $\phi$.
\allowdisplaybreaks
\begin{small}
\begin{align}\label{16_odd}
&o_1(\phi)\,\epsilon_{\zeta \eta \theta \kappa}R_{\alpha \beta \gamma \delta}  R^{\alpha \beta\zeta \eta} R^{\gamma \delta\theta \kappa},&&\!\!\!\!o_2(\phi)\,\epsilon_{\gamma \delta \zeta\eta}\nabla_{\theta} R_{\alpha \beta}{}^{\gamma \delta}\nabla^{\theta} R^{\alpha \beta\zeta\eta} ,&&\!\!\!\!o_3(\phi)\,\epsilon_{\gamma \delta \zeta \eta}R_{\alpha \beta}{}^{\gamma \delta} R^{\alpha \beta\zeta \eta} R,\nn\\
&o_4(\phi)\,\epsilon_{\gamma \delta\zeta \eta}R_{\alpha \beta}{}^{\gamma \delta} R^{\alpha \zeta} R^{\beta \eta} ,&&\!\!\!\!o_5(\phi)\,\epsilon_{\gamma \delta\zeta \eta} R_{\alpha \beta}{}^{\gamma \delta} \nabla^{\eta} \nabla^{\beta} R^{\alpha \zeta},&&\!\!\!\!o_6(\phi)\,\epsilon_{\beta \gamma \delta \zeta} \nabla^\gamma R_{\alpha}{}^\beta\nabla^{\zeta} R^{\alpha\delta},\nn\\[-3mm]
&\quad\,\mathclap{\rule{1cm}{0.4pt}}\nn\\[-2mm]
&o_7(\phi)\,\epsilon_{\gamma\delta\zeta\eta}R_{\alpha\beta}{}^{\gamma\delta} R^{\alpha\beta\zeta\eta}  \nabla_{\theta} \phi \nabla^{\theta} \phi,&&\!\!\!\!o_8(\phi)\,\epsilon_{\gamma \delta \zeta \eta}R_{\alpha\beta}{}^{\gamma \delta} R^{\alpha \beta\zeta \eta} \,\bo\phi,&&\!\!\!\!o_9(\phi)\,\epsilon_{\theta \delta \zeta\eta} R_{\alpha\beta}{}^{\gamma\delta} \nabla_{\gamma} R^{\alpha\beta\zeta\eta}\nabla^{\theta} \phi,\nn\\
&o_{10}(\phi)\,\epsilon_{\gamma \delta\zeta \eta}R_{\alpha \beta}{}^{\gamma \delta} R^{\beta \eta}  \nabla^{\alpha} \phi \nabla^{\zeta} \phi,&&\!\!\!\!o_{11}(\phi)\,\epsilon_{\gamma \delta\zeta \eta}R_{\alpha \beta}{}^{\gamma \delta} R^{\alpha \zeta}  \nabla^{\eta} \nabla^{\beta} \phi,&&\!\!\!\!o_{12}(\phi)\,\epsilon_{\gamma \delta\zeta \eta} R_{\alpha \beta}{}^{\gamma \delta} \nabla^{\beta} R^{\alpha \zeta}\nabla^{\eta} \phi,\nn\\
&o_{13}(\phi)\,\epsilon_{\gamma \delta\zeta \eta} R_{\alpha \beta}{}^{\gamma \delta}\,\nabla^{\alpha} \phi \nabla^{\zeta} \phi \nabla^{\eta} \nabla^{\beta} \phi,&&\!\!\!\!o_{14}(\phi)\,\epsilon_{\gamma \delta\zeta \eta} R_{\alpha \beta}{}^{\gamma \delta} \nabla^{\zeta} \nabla^{\alpha} \phi \nabla^{\eta} \nabla^{\beta} \phi,&&\!\!\!\!o_{15}(\phi)\,\epsilon_{\beta \gamma \delta\zeta} R_\alpha{}^\beta \nabla^{\delta}R^{\alpha\gamma}\nabla^{\zeta} \phi,\nn\\
&o_{16}(\phi)\,\epsilon_{ \beta \gamma\delta \zeta}\nabla^{\gamma} R_\alpha{}^\beta\, \nabla^{\delta} \phi \nabla^{\zeta} \nabla^{\alpha} \phi.
\end{align}
\end{small}
After integrating by parts, nine independent terms remain. Requiring that the derivatives act with the lowest possible order on the scalar or tensor fields, we obtain the following set of terms:
\allowdisplaybreaks
\begin{small}
\begin{align}\label{odd_offshell}
&o_1(\phi)\,\epsilon_{\zeta \eta \theta \kappa}R_{\alpha \beta \gamma \delta}  R^{\alpha \beta\zeta \eta} R^{\gamma \delta\theta \kappa},&&\!\!\!\!o_3(\phi)\,\epsilon_{\gamma \delta \zeta \eta}R_{\alpha \beta}{}^{\gamma \delta} R^{\alpha \beta\zeta \eta} R,&&\!\!\!\!o_4(\phi)\,\epsilon_{\gamma \delta\zeta \eta}R_{\alpha \beta}{}^{\gamma \delta} R^{\alpha \zeta} R^{\beta \eta} ,\nn\\[-3mm]
&\quad\,\mathclap{\rule{1cm}{0.4pt}}\nn\\[-2mm]
&o_7(\phi)\,\epsilon_{\gamma\delta\zeta\eta}R_{\alpha\beta}{}^{\gamma\delta} R^{\alpha\beta\zeta\eta}  \nabla_{\theta} \phi \nabla^{\theta} \phi,&&\!\!\!\!o_8(\phi)\,\epsilon_{\gamma \delta \zeta \eta}R_{\alpha\beta}{}^{\gamma \delta} R^{\alpha \beta\zeta \eta} \,\bo\phi,&&\!\!\!\!o_{10}(\phi)\,\epsilon_{\gamma \delta\zeta \eta}R_{\alpha \beta}{}^{\gamma \delta} R^{\beta \eta}  \nabla^{\alpha} \phi \nabla^{\zeta} \phi,\nn\\
&o_{11}(\phi)\,\epsilon_{\gamma \delta\zeta \eta}R_{\alpha \beta}{}^{\gamma \delta} R^{\alpha \zeta}  \nabla^{\eta} \nabla^{\beta} \phi,&&\!\!\!\!o_{13}(\phi)\,\epsilon_{\gamma \delta\zeta \eta} R_{\alpha \beta}{}^{\gamma \delta}\,\nabla^{\alpha} \phi \nabla^{\zeta} \phi \nabla^{\eta} \nabla^{\beta} \phi,&&\!\!\!\!o_{15}(\phi)\,\epsilon_{\beta \gamma \delta\zeta} R_\alpha{}^\beta \nabla^{\delta}R^{\alpha\gamma}\nabla^{\zeta} \phi.
\end{align}
\end{small}
Adding terms \eqref{even_offshell} and \eqref{odd_offshell} gives the six-derivative off-shell Lagrangian. Taking this off-shell Lagrangian on-shell gives us Lagrangian \eqref{6der_onshell}.	


	\bibliographystyle{utphys}
	\bibliography{bib1.bib}
\end{document}